\documentclass[a4paper,11pt]{article}
\usepackage{jheppub}
\usepackage[utf8]{inputenc}

\usepackage{dsfont}
\usepackage{simplewick}

\usepackage{mathtools,amsmath,amssymb}
\usepackage{graphicx}

\usepackage{lmodern}
\usepackage[T1]{fontenc}
\usepackage{microtype}

\usepackage{hyperref}

\usepackage{xspace}

\usepackage{array} 

\usepackage{picture}
\usepackage{calc}

\usepackage{subfigure}

\usepackage{pgf}
\usepackage{tikz}
\usetikzlibrary{shapes,arrows,automata,positioning}
\usepackage{tikz-3dplot}

\newcommand{\YM}{{\mathrm{\scriptscriptstyle YM}}}

\usepackage{xspace}
\usepackage{tikz}
\usetikzlibrary{shapes}
\usetikzlibrary{decorations.markings}

\tikzset{->-/.style={decoration={
  markings,
  mark=at position .5 with {\arrow{>}}},postaction={decorate}}}

\usepackage{placeins}

\DeclareMathOperator{\phaneq}{\phantom{{}=}}

\newcommand{\complicated}{\text{com.}}

\newcommand{\half}{{\tfrac{1}{2}}}

\newcommand{\mc}[1]{\mathcal{#1}}

\newcommand{\easy}{E}
\newcommand{\gammaE}{\gamma_{\text{E}}}
\newcommand{\e}{\operatorname{e}}
\newcommand{\de}{\operatorname{d}\!}

\newcommand{\peps}{\varepsilon}
\newcommand{\teps}{\epsilon}

\newcommand{\mds}[1]{\mathds{#1}}

\newcommand{\eqndot}{\, . }
\newcommand{\eqncom}{\, , }

\DeclareMathOperator{\tr}{tr}

\renewcommand{\parallel}{\upharpoonleft\!\upharpoonright}
\newcommand{\aparallel}{\upharpoonleft\!\downharpoonright}

\DeclareMathOperator{\idm}{\mathds{1}}

\DeclareMathOperator{\Ad}{Ad}

\DeclareMathOperator{\sgn}{sgn}

\makeatletter
\renewcommand*\env@matrix[1][\arraystretch]{%
  \edef\arraystretch{#1}%
  \hskip -\arraycolsep
  \let\@ifnextchar\new@ifnextchar
  \array{*\c@MaxMatrixCols c}}
\makeatother

\makeatletter
\def\mmatrix{\@ifnextchar[{\@amwith}{\@mwithout}}
\def\@amwith[#1]{\@ifnextchar[{\@mwwith[#1]}{\@mwith[#1]}}
\def\@mwwith[#1][#2]#3{
\begingroup\setlength{\arraycolsep}{#2pt}
    \begin{pmatrix}[#1]
        #3
    \end{pmatrix}
\endgroup
}
\def\@mwith[#1]#2{
    \begin{pmatrix}[#1]
        #2
    \end{pmatrix}
}
\def\@mwithout#1{
    \begin{pmatrix}
        #1
    \end{pmatrix}
}
\makeatother

\makeatletter
\DeclareRobustCommand*{\bfseries}{%
  \not@math@alphabet\bfseries\mathbf
  \fontseries\bfdefault\selectfont
  \boldmath
}

\makeatother

\usepackage{lipsum}
\makeatletter
\if@titlepage
  \renewenvironment{abstract}{%
      \titlepage
      \null\vfil
      \@beginparpenalty\@lowpenalty
      \begin{center}%
        \bfseries \abstractname
        \@endparpenalty\@M
      \end{center}}%
     {\par\vfil\null\endtitlepage}
\else
  \renewenvironment{abstract}{%
      \if@twocolumn
        \section*{\abstractname}%
      \else
        \small
        \begin{center}%
          {\bfseries \abstractname\vspace{-.5em}\vspace{\z@}}%
        \end{center}%
        \quotation
      \fi}
      {\if@twocolumn\else\endquotation\fi}
\fi
\makeatother

\newcommand{\SU}[1]{\operatorname{SU}(#1)}
\newcommand{\SO}[1]{\operatorname{SO}(#1)}

\renewcommand{\digamma}{\Psi}
        
\DeclareMathOperator{\cder}{D}
\newcommand{\scal}{\phi}
\newcommand{\scalc}{\scal^{\text{cl}}}
\newcommand{\scalq}{\tilde{\scal}}
\newcommand{\ferm}{\psi}
\newcommand{\aferm}{\bar{\ferm}}
\newcommand{\tenDferm}{\Psi}
\newcommand{\atenDferm}{\bar{\tenDferm}}
\newcommand{\comm}[2]{[#1,#2]}

\title{A Quantum Check of AdS/dCFT}
\author{Isak Buhl-Mortensen, Marius de Leeuw, Asger C. Ipsen, Charlotte Kristjansen and Matthias Wilhelm}

\begin{document}

\begingroup\parindent0pt
\begin{flushright}\footnotesize
\end{flushright}
\vspace*{4em}
\centering
\begingroup\LARGE
\bf
A Quantum Check of AdS/dCFT 
\par\endgroup
\vspace{2.5em}
\begingroup\large{\bf Isak Buhl-Mortensen, Marius de Leeuw, Asger C.\ Ipsen, \\Charlotte Kristjansen and Matthias Wilhelm}
\par\endgroup
\vspace{1em}
\begingroup\itshape
Niels Bohr Institute, Copenhagen University,\\
Blegdamsvej 17, 2100 Copenhagen \O{}, Denmark\\

\par\endgroup
\vspace{1em}
\begingroup\ttfamily
buhlmort@nbi.ku.dk, deleeuwm@nbi.ku.dk, asgercro@nbi.ku.dk, kristjan@nbi.ku.dk, matthias.wilhelm@nbi.ku.dk \\
\par\endgroup
\vspace{2.5em}
\endgroup

\begin{abstract}
\noindent
We build the framework for performing loop computations in the defect version of ${\cal N}=4$ super Yang-Mills theory which 
is dual to the probe D5-D3 brane system with background gauge-field flux. In this dCFT, a codimension-one defect separates
two regions of space-time with different ranks of the gauge group and three of the scalar fields acquire 
non-vanishing and space-time-dependent vacuum expectation values. 
The latter leads to a highly non-trivial mass mixing problem between different colour and flavour components, which we solve using fuzzy-sphere coordinates. 
Furthermore, the resulting space-time dependence of the theory's Minkowski space propagators is handled by reformulating these as propagators in an effective $AdS_4$.
Subsequently, we initiate the computation of quantum corrections.
The  one-loop correction to the one-point function of any local gauge-invariant scalar operator is shown to receive contributions from only two Feynman diagrams. We regulate these diagrams using dimensional reduction, 
finding that one of the two diagrams vanishes, and discuss the procedure for calculating the one-point function of a generic operator from the $\SU2$ subsector.
Finally, we explicitly evaluate  the one-loop correction to the one-point function of the BPS vacuum state, finding perfect agreement with an earlier string-theory prediction. 
This constitutes a highly non-trivial test of the gauge-gravity duality in a situation where both supersymmetry and conformal symmetry are partially broken.
\end{abstract}


\thispagestyle{empty}

\newpage
\hrule
\tableofcontents
\afterTocSpace
\hrule
\afterTocRuleSpace

\section{Introduction}
\label{sec: introduction}
Defect conformal field theories (dCFTs) with holographic duals constitute an interesting new arena for 
precision tests of the AdS/CFT correspondence~\cite{Maldacena:1997re} 
and for the search for integrable structures~\cite{Beisert:2010jr}.
Moreover, for such quantum field theories new types of correlation
functions come into play. For instance, fields living on the defect can mix with bulk fields and two-point functions of bulk fields with unequal conformal dimensions need not vanish~\cite{Cardy:1984bb}.
Further interesting features emerge if one considers set-ups where some of the bulk fields acquire a vacuum expectation value (vev), in which case the theory can have non-vanishing one-point functions already
at tree level~\cite{Cardy:1984bb,DeWolfe:2001pq}.
The study of one-point functions is a natural first step when entering the realm of dCFTs. Tree-level studies carried out  within the AdS/dCFT framework show that one-point functions, interestingly, have many features in common with three-point functions of the standard AdS/CFT set-up, e.g.\ determinant-based expressions, integrable structure and an accessible strong-coupling limit~\cite{deLeeuw:2015hxa,Buhl-Mortensen:2015gfd,deLeeuw:2016umh}.

In the present paper, we shall develop the necessary tools to go beyond tree-level computations in certain dCFTs with vevs 
and with holographic duals, an endeavour which will make possible the extraction of large amounts of new data from these theories  as well as the initiation of  new directions of study. We already briefly presented one example of a one-loop analysis in such a dCFT in the letter~\cite{Buhl-Mortensen:2016pxs}, 
where we calculated the one-loop correction to the one-point function of a chiral primary and  compared it to the result of a string-theory computation in a certain double-scaling limit,
finding exact agreement. Here, we present the derivations which made
the field-theoretic part of that  computation possible, give the details of the computation and extend these results to finite $N$ as well as to general single-trace operators built out of scalar fields.

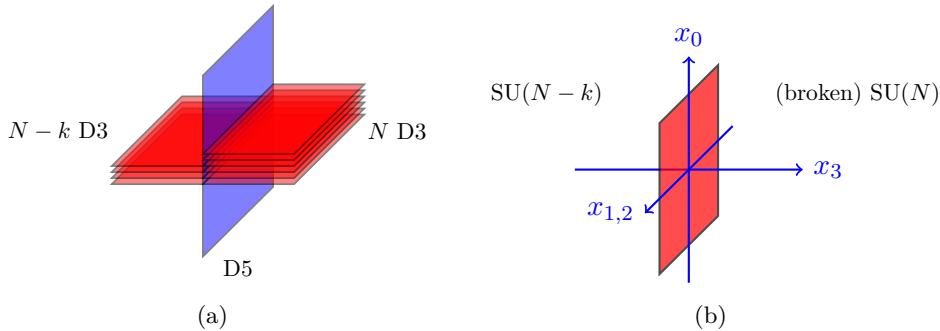
\begin{figure}[htbp]
 \centering
 \subfigure[]{
\centering
\scalebox{0.8}{
\begin{tikzpicture}
	[	D3/.style={opacity=.5,thick,fill=red},
		D5/.style={opacity=.5,thick,fill=blue}]
	\foreach \x in {-0.3,-0.2,...,0.1}
		{
	\draw[D3] (-1.5,\x,-1.5) -- (0,\x,-1.5) -- (0,\x,1.5) -- (-1.5,\x,1.5) -- cycle;
}	
	\draw[D5] (0,-1.5,-1.5) -- (0,+1.5,-1.5) -- (0,1.5,1.5) -- (0,-1.5,1.5) -- cycle;
	\foreach \x in {-0.3,-0.2,...,0.3}
		{
	\draw[D3] (0,\x,-1.5) -- (1.5,\x,-1.5) -- (1.5,\x,1.5) -- (0,\x,1.5) -- cycle;
}
	\node[anchor=west] at (2,0,0) {$N$ D3};
	\node[anchor=east] at (-2,0,0) {$N-k$ D3};
	\node[anchor=north] at (0,-2,0) {D5};
\end{tikzpicture}
}
\label{subfig: branes}
} 
 \subfigure[]{
\centering
\scalebox{1}{
\begin{tikzpicture}
	[	axis/.style={->,blue,thick},
		axisline/.style={blue,thick},
		cube/.style={opacity=.7, thick,fill=red}]
	\draw[axisline] (-1.5,0,0) -- (0,0,0) node[anchor=west]{};	
	\draw[cube] (0,-1,-1) -- (0,1,-1) -- (0,1,1) -- (0,-1,1) -- cycle;
	\draw[axis] (0,0,0) -- (1.5,0,0) node[anchor=west]{$x_3$};
	\draw[axis] (0,-1.5,0) -- (0,1.5,0) node[anchor=south]{$x_0$};
	\draw[axis] (0,0,-1.5) -- (0,0,1.5) node[anchor=east]{$x_{1,2}$};	
	\node[anchor=east] at (-1,1,0) {\scalebox{0.8}{$\SU{N-k}$}};
	\node[anchor=west] at (1,1,0) {\scalebox{0.8}{(broken) $\SU{N}$}};
\end{tikzpicture}
}
\label{subfig: gauge theory}
}  
\caption{Illustration of the set-up: \subref{subfig: branes} $k$ of the $N$ D3 branes get dissolved in the D5 probe brane \subref{subfig: gauge theory} the rank of the gauge group differs on the two sides of the defect. \label{dCFTset-up}
}
\end{figure}

 The dCFT we are going to consider consists of ${\cal N}=4$ super Yang-Mills ($\mathcal{N}=4$ SYM) theory with a codimension-one defect inserted at  $x_3=0$~\cite{DeWolfe:2001pq}. Three of the scalar fields of the theory are assigned specific, $x_3$-dependent vevs on one side of the defect, $x_3>0$, while all classical fields vanish for $x_3<0$.  
  This Higgsing  results in a highly non-trivial mass mixing problem where different colour components for both bosonic and fermionic fields mix with each other and where in addition one space-time component of the gauge field mixes with the scalars.  Moreover, all mass terms become $x_3$-dependent. The
motivation for this particular Higgsing comes from the string-theory set-up, where the vevs represent the so-called fuzzy-funnel
solution of the probe D5-D3 brane system where the probe-D5 brane is embedded in $AdS_5\times S^5$ so that it shares
three dimensions (the defect) with the $N$ D3 branes. 
More precisely, the geometry of the D5 brane is 
$AdS_4\times S^2$ and a certain background gauge field has a non-vanishing flux, $k$, on 
$S^2$~meaning that $k$ out of the $N$ D3 branes get dissolved in the D5 brane \cite{Karch:2000gx,Nahm:1979yw,Diaconescu:1996rk,Constable:1999ac}.  
On the gauge theory side, the parameter $k$ appears as the difference in rank of the 
gauge group on the two sides of the defect, cf.\ figure~\ref{dCFTset-up}.

Due to the Higgsing, the theory  has non-vanishing one-point functions already at tree level. Tree-level one-point functions
of chiral primaries were calculated for this particular theory in~\cite{Nagasaki:2011ue} as well as in a closely related one in~\cite{Kristjansen:2012tn}, and a match with a string-theory computation was found at the leading order in a certain double-scaling limit.  Moreover, making use
of the tools of integrability, it was possible to derive a closed expression of determinant form for the tree-level
one-point functions of non-protected operators belonging to an $\SU2$ subsector of ${\cal N}=4$ 
SYM theory~\cite{deLeeuw:2015hxa, Buhl-Mortensen:2015gfd}.  An empirically
based proposal for how to extend this to an $\SU3$ sector likewise exists~\cite{deLeeuw:2016umh}.

Due to the mass mixing problem, going beyond tree-level for the Higgsed theory is considerably more complicated than for ${\cal N}=4$ SYM theory itself.  It turns out, however, that the language of  fuzzy-sphere coordinates is tailored for the diagonalisation of
the mass matrix. 
In these coordinates, the mixing problem can literally be viewed as the spin-orbit interaction of the
hydrogen atom of the 21st century, $\mathcal{N}=4$ SYM theory.
Furthermore, it is possible to avoid the space-time dependence of the masses by formulating the propagators
in an effective $AdS_4$ space.  The radial coordinate of this $AdS_4$ space is $x_3$, the  coordinate perpendicular to the defect, and the defect itself plays the role of the $AdS_4$ boundary. 
With these steps accomplished, the theory is in principle amenable to the standard program of perturbation theory.  
We show that the one-loop correction to any (single-trace) operator built from scalars obtains contributions from only two Feynman diagrams 
and we calculate these using dimensional regularisation in combination with dimensional reduction carefully adjusted to respect the symmetries of the
present set-up. One of the two relevant Feynman diagrams corresponds to  the one-loop correction to the vevs of the scalars and cancels exactly.

We discuss in some depth the computation
of one-loop corrections to one-point functions in the $\SU2$ subsector and, in particular, 
we present the details of the calculation of the planar correction to the one-point function of the BMN vacuum state, the result of which we presented in the letter~\cite{Buhl-Mortensen:2016pxs}.  Here, we adress the finite-$N$ case as well. 

The first step of our perturbative calculation consists in expanding the SYM action around the classical fields
and fixing an appropriate gauge. This step is carried out in section~\ref{sec:action}. Section~\ref{massdiagonalisation} is devoted
to the resolution of the mass mixing problem. 
First, we rewrite the mass term in terms of irreducible $\SU2$ representations in flavour space. Then, we explicitly construct the eigenstates via fuzzy-sphere coordinates and a Clebsch-Gordan decomposition.
The section closes with a table of the resulting spectrum
of the theory, cf. page~\pageref{spectrum}.  As all mass terms carry space-time dependence, being proportional to 
$1/x_3$ for fermions and $1/(x_3)^2$ for bosons,
the propagators of the theory are not of standard Minkowskian type. We show in section~\ref{Propagators} that the propagators can be viewed as
standard propagators of $AdS_4$ instead. 
Moreover, we translate the propagators in the mass eigenbasis to the flavour and colour basis.
We discuss the dimensional regularisation of the occurring integrals as well as dimensional reduction in section \ref{dimreg}.
Section~\ref{oneloop} deals with the computation of  one-loop corrections 
to  one-point functions of scalar operators, first in general, subsequently for operators belonging to the $\SU{2}$ subsector
and finally for the BMN vacuum state.  
We are mainly working in the planar limit but include a number of finite $N$ results as well. 
The computation of the one-loop correction to the vevs of the scalar fields, which is required for 
 the analysis of this section, is relegated to
appendix \ref{app: one-loop vevs}.  Section~\ref{stringtheory} is devoted to the comparison to string theory and finally section~\ref{sec: conclusion and outlook}
contains
 a conclusion and outlook, where we discuss a number of other interesting quantum computations  for dCFTs which our
work makes feasible. 
Five appendices provide details on various aspects of our work: the irreducible $\SU2$ representations (\ref{representationmatrices}), the fuzzy-sphere coordinates (\ref{fuzzy}), our conventions for the ten-dimensional gamma matrices (\ref{app: 10Dto4D}), the aforementioned calculation of the vevs of the scalars (\ref{app: one-loop vevs}) and the alternative Hadamard and zeta-function regularisation (\ref{Hadamard}).

\section{The action}
\label{sec:action}

The action of the dCFT is the sum of the usual ${\cal N}=4$ SYM action in the bulk and an action describing the self-interactions of a 3D hypermultiplet of fundamental fields living on the defect 
and their couplings to the fields of ${\cal N}=4$ SYM theory:
\begin{equation}
S=S_{{\cal N}=4}+S_{D=3}\eqndot
\end{equation}
The defect fields will turn out to play no role at the loop order we consider.
We will use the action of $\mathcal{N}=4$ SYM theory in the following form
\begin{multline}
 S_{{\cal N}=4}=\frac{2}{g_\YM^2}\int \de^4x\tr\Bigg[ -\frac{1}{4}F_{\mu\nu}F^{\mu\nu}-\frac{1}{2}\cder_\mu\scal_i\cder^\mu\scal_i+\frac{i}{2}\atenDferm\Gamma^\mu\cder_\mu\tenDferm \\+\frac{1}{2}\atenDferm\tilde\Gamma^i\comm{\scal_i}{\tenDferm}+\frac{1}{4}\comm{\scal_i}{\scal_j}\comm{\scal_i}{\scal_j}\Bigg]\eqncom
 \label{eq:SYM-action}
\end{multline}
where
\begin{equation}
 \begin{aligned}
  F_{\mu\nu}&=\partial_\mu A_\nu-\partial_\nu A_\mu-i\comm{A_\mu}{A_\nu}\eqncom\\
  \cder_\mu\scal_i&=\partial_\mu\scal_i-i\comm{A_\mu}{\scal_i}\eqncom \hspace{0.5cm}
  \cder_\mu\tenDferm=\partial_\mu\tenDferm-i\comm{A_\mu}{\tenDferm}\eqndot
 \end{aligned}
\end{equation}
Here, the field $\tenDferm$ is a ten-dimensional Majorana-Weyl fermion and  $\{\Gamma^{\mu},\tilde\Gamma^i\}$ are the corresponding ten-dimensional gamma matrices, which we explicitly give in appendix \ref{app: 10Dto4D}. The ranges of the indices are $\mu,\nu=0,1,2,3$ and $i,j=1,2,3,4,5,6$. We are using a mostly-plus convention for the metric.

We wish to expand the fields around the classical solution
\begin{equation} \label{phicl}
 \langle\scal_i\rangle_{\text{tree}}=\scalc_i=-\frac{1}{x_3} t_i\oplus 0_{(N-k)\times(N-k)}\eqncom
\end{equation}
where $i=1,2,3$ and the $t_i$ constitute a $k$-dimensional irreducible representation of the Lie algebra $\SU{2}$;  expressions for the representation matrices in our conventions can be found in appendix~\ref{representationmatrices}. 
All other classical fields vanish. This solution is 
the gauge-theory dual of the fuzzy-funnel solution of the probe D5-D3 brane set-up~\cite{Constable:1999ac}.

We expand the action around the classical solution, writing
\begin{equation}\label{fieldexpansion}
 \scal_i=\scalc_i+\scalq_i\eqncom
\end{equation}
where $\scalc_i$ denotes the classical part and $\scalq_i$ the quantum part.
Terms which upon expansion do not depend on any quantum fields can be ignored as can all terms linear in 
the quantum fields as these should vanish by the equations of motion. This latter fact can also be checked explicitly.

\subsection{Gauge fixing}

As usual, we have to fix a gauge in order to perform calculations.
Moreover, we notice that the expansion of the gauge-kinetic term of the scalar contains
\begin{equation}
 i\comm{A_\mu}{\scalc_i}\partial^\mu\scalq_i\eqncom
\end{equation}
which would lead to complications in computing the propagators.
Hence, we want to cancel this term while fixing the gauge.
Following \cite{Alday:2009zm}, this can be achieved by adding the gauge-fixing term
\begin{equation}
 -\frac{1}{2}\tr(G^2) \quad \text{with} \quad  G=\partial_\mu A^\mu+i\comm{\scalq_i}{\scalc_i}
\end{equation}
to the action.
The price for doing this is a massive ghost field that couples to the scalars.

Explicitly, we add to the action \eqref{eq:SYM-action} the BRST exact term
\begin{equation}
  S_{\text{gh}}
    = \frac{2}{g_\YM^2}\int \de^4 x \tr\left[-s\left(\bar{c}(\partial_\mu A^\mu -i[\scalc_i,\scalq_i]) 
         + \frac{1}{2}\bar{c}B\right)\right]\eqncom
\end{equation}
where $s$ is the BRST variation defined by
\begin{equation}
\begin{aligned}
  sA_\mu &= \cder_\mu c = \partial_\mu c -i[A_\mu,c]\eqncom &
    s\phi_i &= -i[\phi_i,c]\eqncom &
    s\Psi &= i\{\Psi,c\}\eqncom\\
  sc &= ic^2\eqncom &
    s\bar{c} &= -B\eqncom &
    sB &= 0\eqndot
\end{aligned}
\end{equation}
One can check that with this definition $s^2 = 0$.
The ghosts $c,\bar{c}$ are fermionic (Lorentz) scalars, while the auxiliary
field $B$ is a bosonic scalar. The BRST variation only acts on the quantum part of $\phi_i$,
i.e.\
\begin{equation}
  s\scalc_i = 0\eqncom\qquad
  s\scalq_i = -i[\scalc_i+\scalq_i,c]\eqndot
\end{equation}
We now find, noting that moving $s$ past a fermion introduces a sign,
\begin{equation}
  S_{\text{gh}}
    = \frac{2}{g_\YM^2}\int \de^4 x \tr\left[\bar{c}(\partial_\mu \cder^\mu c 
       -[\scalc_i,[\scalc_i+\scalq_i,c]])
       +B(\partial_\mu A^\mu -i[\scalc_i,\scalq_i])
         + \frac{1}{2}B^2\right]\eqndot
\end{equation}
Since $B$ is not dynamical, we can immediately integrate it out; its equation of motion  
is $B = -\partial_\mu A^\mu + i[\scalc_i,\scalq_i]$.
After rearranging the result a bit, this yields
\begin{multline}
  S_{\text{gh}}
    = \frac{2}{g_\YM^2}\int \de^4 x \tr\biggl[\bar{c}(\partial_\mu \cder^\mu c 
       -[\scalc_i,[\scalc_i+\scalq_i,c]])
       -\frac{1}{2}(\partial_\mu A^\mu)^2
       +i[A^\mu,\scalq_i]\partial_\mu\scalc_i\\
       +i[A^\mu,\partial_\mu\scalq_i]\scalc_i
       +\frac{1}{2}[\scalc_i,\scalq_i]^2\biggr]\eqndot
\end{multline}
We note that this cancels the unwanted mixing between $A_\mu$ and $\partial_\mu\scalq_i$,
as mentioned above. 
We also see that the kinetic term for the gluons is changed to
\begin{equation}
  -\frac{1}{4}(\partial_\mu A_\nu-\partial_\nu A_\mu)^2
     -\frac{1}{2}(\partial_\mu A^\mu)^2
  = \frac{1}{2} A_\mu\partial_\nu\partial^\nu A^\mu\eqncom
\end{equation}
which is invertible and diagonal in the Lorentz index. Notice that for $\scalc_i=0$ our gauge choice reduces to 
Feynman gauge.

\subsection{The expanded action}

We can write the gauge-fixed action as
\begin{equation}
  S_{{\cal N}=4} + S_{\text{gh}}
    = S_{\text{kin}} + S_{\text{m,b}} + S_{\text{m,f}}
      + S_{\text{cubic}} + S_{\text{quartic}}\eqndot
\end{equation}
The Gau\ss{}ian part consists of the kinetic terms
\begin{equation}
  S_{\text{kin}}
    = \frac{2}{g_\YM^2}\int \de^4 x \tr\left[
        \frac{1}{2}A_\mu \partial_\nu\partial^\nu A^\mu
        +\frac{1}{2}\scalq_i \partial_\nu\partial^\nu \scalq_i
        +\frac{i}{2}\aferm\gamma^\mu\partial_\mu\ferm
        +\bar{c}\partial_\mu\partial^\mu c\right]\eqncom
\end{equation}
the bosonic mass terms
\begin{multline}
  S_{\text{m,b}}
    = \frac{2}{g_\YM^2}\int \de^4 x \tr\biggl[
        \frac{1}{2}\comm{\scalc_i}{\scalc_j}\comm{\scalq_i}{\scalq_j}
        +\frac{1}{2}\comm{\scalc_i}{\scalq_j}\comm{\scalc_i}{\scalq_j}
        +\frac{1}{2}\comm{\scalc_i}{\scalq_j}\comm{\scalq_i}{\scalc_j}\\
        +\frac{1}{2}[\scalc_i,\scalq_i][\scalc_j,\scalq_j]
        +\frac{1}{2}[A_\mu,\scalc_i][A^\mu,\scalc_i]
        +2i[A^\mu,\scalq_i]\partial_\mu\scalc_i\biggr]\eqncom
  \label{eq:bosonic-mass}
\end{multline}
and the fermionic mass terms
\begin{equation}
  S_{\text{m,f}}
    = \frac{2}{g_\YM^2}\int \de^4 x \tr\left[
        \frac{1}{2}\aferm G^i[\scalc_i,\ferm]
        -\bar{c}[\scalc_i,[\scalc_i,c]]\right]\eqncom
  \label{eq:fermionic-mass}
\end{equation}
where we have reduced the ten-dimensional Majorana-Weyl fermion to four four-dimensional Majorana
fermions $\psi_j$, $j=1,2,3,4$, as explained in appendix~\ref{app: 10Dto4D}, and the $4\times4$ matrices $G^i$ that describe their coupling to the scalars are given in \eqref{eq:G123}. 
The interaction is given by the cubic vertices
\begin{multline}
  S_{\text{cubic}}
    = \frac{2}{g_\YM^2}\int \de^4 x \tr\biggl[
        i[A^\mu,A^\nu]\partial_\mu A_\nu
        +[\scalc_i,\scalq_j][\scalq_i,\scalq_j]
        +i[A^\mu,\scalq_i]\partial_\mu\scalq_i
        +[A_\mu,\scalc_i][A^\mu,\scalq_i]\\
        +\frac{1}{2}\aferm\gamma^\mu[A_\mu,\ferm]
        +\sum_{i=1}^3\frac{1}{2}\aferm G^i[\scalq_i,\ferm]
        +\sum_{i=4}^6\frac{1}{2}\aferm G^i[\scalq_i,\gamma_5\ferm]
        +i(\partial_\mu\bar{c})[A^\mu,c]
        -\bar{c}[\scalc_i,[\scalq_i,c]]\biggr]
        \label{eq:cubic}
\end{multline}
and the quartic vertices
\begin{equation}
  S_{\text{quartic}}
    = \frac{2}{g_\YM^2}\int \de^4 x \tr\left[
        \frac{1}{4}[A_\mu,A_\nu][A^\mu,A^\nu]
        +\frac{1}{2}[A_\mu,\scalq_i][A^\mu,\scalq_i]
        +\frac{1}{4}[\scalq_i,\scalq_j][\scalq_i,\scalq_j]\right]\eqndot
        \label{eq:quartic}
\end{equation}
We shall see below that $S_{\text{quartic}}$ is not relevant for the one-loop corrections in this article.
In the remainder of the paper, we will work in Euclidean signature.

\section{The mass matrix \label{massdiagonalisation}}

The mass terms of the action~(\ref{eq:bosonic-mass}) and~(\ref{eq:fermionic-mass}) involve mixing between fields of different flavour as well as mixing between colour components of the same field. 
To prepare for perturbative calculations of correlation functions, we first have to solve this highly non-trivial mixing problem. 
Notice that the mass terms are also unconventional in the sense that they depend via the classical fields on the distance $x_3$ to the defect. 
This $x_3$-dependence renders some of the traditional tools of quantum field theory in Minkowski space inapplicable. 
We will show how to deal with this issue by trading $x_3$-dependent 4d Minkowski space propagators for $x_3$-independent propagators in $AdS_4$ in the next section.

Let us now diagonalise the mass matrix. First, in subsection~\ref{subsec: rewriting} we rewrite the mass terms in close analogy to the spin-orbital interaction of the hydrogen atom, so that they are easy to diagonalise. 
Subsequently, in subsection~\ref{diagonalisation} we
explicitly carry out the diagonalisation and read off the spectrum including its degeneracies.
We summarise our results on the spectrum in subsection \ref{subsec: summary}.

\subsection{Rewriting of the mass terms}
\label{subsec: rewriting}

For a sub-set of the fields, the mass terms are diagonal in the flavor index (but not in the colour index)
and we denote the corresponding fields as easy fields. Accordingly, the remaining fields are denoted as complicated fields.
The easy fields consist of the three scalars $\phi_4,\phi_5,\phi_6$, the three gauge fields $A_0,A_1,A_2$ and
the ghost $c$.

For the easy fields, say $A_0$ for concreteness, the mass term is proportional to
\begin{equation}
  \tr([t_i,A_0][t_i,A_0])
    = -\tr(A_0[t_i,[t_i,A_0]])
    = -\tr(A_0 L^2 A_0)\eqncom
\end{equation}
where 
\begin{equation}
  L_i = \Ad(t_i)\eqncom\qquad L^2 = L_i L_i
  \label{eq:fuzzy-L}
\end{equation}
are satisfying the well-known commutation relations of angular momenta:
\begin{equation}
 [L_i,L_j]=i\teps_{ijk}L_k\eqndot
\end{equation}
The operator $L^2$ is the Laplacian on the so-called fuzzy sphere.
The field $A_0$ transforms in a -- in general reducible -- representation of the Lie algebra $\SU2$.
We will decompose this representation into irreducible representations with definite orbital quantum number $\ell$ and magnetic quantum number $m$ in the next subsection.

The mass term for the complicated bosons, i.e.\ $\phi_1,\phi_2,\phi_3$ and $A_3$, reads
\begin{equation}
  S_{\text{m,cb}}
    = \frac{2}{g_\YM^2}\int \de^4 x \frac{1}{ x_3^2}\tr \left[
        -\frac{1}{2}\scalq_i L^2 \scalq_i -  \frac{1}{2}A_3 L^2 A_3
        + i\teps_{ijk} \scalq_i L_j \scalq_k
        + i\scalq_i L_i A_3 - i A_3 L_i \scalq_i \right]\eqncom
\end{equation}
where $i=1,2,3$.
We can write this in the more suggestive way 
\begin{equation}
  S_{\text{m,cb}}
    = \frac{2}{g_\YM^2}\int \de^4 x \frac{1}{ x_3^2}\tr \left[C^T(
        -\frac{1}{2} L^2 + 2 S_i L_i) C\right]\eqncom
\end{equation}
where we have introduced the combined field
\begin{equation}
  C = \begin{pmatrix} \scalq_1\\ \scalq_2\\ \scalq_3\\ A_3 \end{pmatrix}\eqncom
\end{equation}
and where the matrices $S_i$ acting on the `flavour' index of $C$ are given by
\begin{equation}
\label{eq:S-matrices}
  S_1 = -\frac{1}{2}\begin{pmatrix}
          0  & \sigma_2 \\
          \sigma_2 & 0 
        \end{pmatrix}\eqncom\quad
  S_2 = \frac{i}{2}
  \begin{pmatrix}
          0 & \idm_2 \\
          -\idm_2 & 0
        \end{pmatrix}\eqncom\quad
  S_3 = \frac{1}{2}
  \begin{pmatrix}
          \sigma_2 & 0 \\
          0 & \sigma_2
        \end{pmatrix}
\end{equation}
with the Pauli matrices
\begin{equation}
\sigma_1 = \mmatrix[1][5]{0 & 1\\ 1 & 0}\eqncom \qquad 
\sigma_2 = \mmatrix[1][3]{0 & -i \\ i & 0 }\eqncom \qquad 
\sigma_3 = \mmatrix[1][3]{1 & 0 \\ 0 & -1}\eqndot
\end{equation}
It is easy to verify that the matrices $S_i$ form a four-dimensional representation of the $\SU2$ Lie algebra:
\begin{equation}
  [S_i,S_j] = i\teps_{ijk}S_k \eqndot
\end{equation}
This representation is reducible and its explicit decomposition into irreducible representations is
\begin{equation}
  U^\dagger S_i U = 
  \begin{pmatrix} \frac{1}{2}\sigma_i &0 \\ 0 &\frac{1}{2}\sigma_i \end{pmatrix}\eqncom
  \quad U = \frac{1}{\sqrt 2}\begin{pmatrix}
               -i &  0 & 0 & i \\
                1 &  0 & 0 & 1 \\
                0 &  i & i & 0 \\
                0 & -1 & 1 & 0
            \end{pmatrix} \eqndot
  \label{eq:Si-similarity}
\end{equation}
The eigenvectors of the irreducible representations are 
\begin{equation}
\label{eq: def top and bottom fields}
  \begin{pmatrix} C_{t,+} \\ C_{t,-} \\ C_{b,+} \\ C_{b,-} \end{pmatrix}
  = U^\dagger C 
  = \frac{1}{\sqrt{2}}\begin{pmatrix}+i\scalq_1+\scalq_2\\
     -i\scalq_3-A_3 \\ -i\scalq_3+A_3\\ 
     -i\scalq_1+\scalq_2 \end{pmatrix}\eqncom
\end{equation}
which have spin $\frac{1}{2}$ and spin magnetic quantum number $\pm\frac{1}{2}$.
It now follows that the complicated boson problem can be solved by 
the usual procedure of adding angular momentum as it occurs in the well-known spin-orbit interaction of the hydrogen atom.
Concretely, we  define the total angular momentum operator
\begin{equation}
 J_i = L_i + \frac 1 2 \sigma_i\eqncom
\end{equation}
and find that
\begin{equation}
  \sigma_i L_i = J^2-L^2-\frac{3}{4}\eqndot
    \label{eq:J-squared}
\end{equation}
We will construct the simultaneous eigenstates of $L^2$, $J^2$ and $J_3$ in the next subsection.

The fermionic mass term is proportional to
\begin{equation}
  \tr[\aferm G^i [t_i,\ferm]]
    = \tr[\aferm G^i L_i \ferm]\eqncom
\end{equation}
where the matrices $G^i$ are given by
\begin{equation} \label{eq:G-matrices}
 G^1 = i
\begin{pmatrix}
0 & -\sigma_3  \\
\sigma_3 & 0 
\end{pmatrix} \eqncom \qquad
G^2 =i
\begin{pmatrix}
 0 & \sigma_1 \\
 -\sigma_1 & 0 \\
\end{pmatrix}
\eqncom \qquad
G^3 = 
\begin{pmatrix}
 \sigma_2 & 0 \\
 0 & \sigma_2 
\end{pmatrix}
\eqndot
\end{equation}
These matrices satisfy the commutation relations
\begin{equation}
 [G^i,G^j]=-2i\teps^{ijk}G^k
\end{equation}
and thus also form a representation of the Lie algebra $\SU2$, at least after a rescaling.
This representation is equally reducible and explicitly reduced as
\begin{equation}
  \label{eq:G123-decomposition}
  \tilde{U}^\dagger G^i \tilde{U} = 
    \begin{pmatrix} -\sigma_i &0 \\ 0 & -\sigma_i \end{pmatrix}\eqncom
  \quad \tilde{U} = \frac{1}{\sqrt 2}\begin{pmatrix}
               0 & -i & -1 &  0 \\
               0 &  1 &  i &  0 \\
              -1 &  0 &  0 &  i \\
               i &  0 &  0 & -1 
            \end{pmatrix} \eqndot
\end{equation}
The eigenvectors of these irreducible representations are 
\begin{equation}
\mmatrix{\psi_{t,+} \\ \psi_{t,-} \\ \psi_{b,+} \\\psi_{b,-}} 
= \tilde{U}^\dagger \psi = \frac{1}{\sqrt{2}}\mmatrix{-\psi_3 - i \psi_4 \\ +\psi_2 + i \psi_1 \\ - \psi_1 - i \psi_2 \\ - \psi_4-i\psi_3 }\eqncom
\end{equation}
which have spin $\frac{1}{2}$ and spin magnetic quantum number $\pm\frac{1}{2}$.
The mixing problem of the fermions can now be solved in complete analogy to the one of the complicated bosons.

To summarise, the complete mass term \eqref{eq:bosonic-mass}, \eqref{eq:fermionic-mass} can be written as
\begin{align}
  \label{eq:new-mass}
  S_{\text{m,b}}+S_{\text{m,f}}
    &= \frac{2}{g_\YM^2}\int \de^4 x \frac{1}{ x_3^2}\tr\biggl[
        -\frac{1}{2}\easy^T L^2\easy       -\bar{c}L^2c
        -\frac{1}{2} C_t^{\dagger}(L^2 - 2 \sigma_i L_i) C_t
        \biggr]
      \\\nonumber
      &\phaneq +\frac{2}{g_\YM^2}\int \de^4 x \frac{1}{ x_3}\tr\biggl[
        \frac{1}{2}\aferm_t \sigma_iL_i\ferm_t
        \biggr]
        +(t\to b)
        \eqncom
\end{align}
where 
\begin{equation}
 \easy= \begin{pmatrix} A_0\\ A_1\\ A_2\\ \scalq_4\\ \scalq_5\\ \scalq_6 \end{pmatrix}\eqndot
\end{equation}
Note that the conjugation here is understood to be outside of the indices, i.e.\
\begin{equation}
 C_t^{\dagger}\equiv(C_t)^\dagger\eqncom \qquad 
 \aferm_t\equiv (\ferm_t)^\dagger\gamma^0\eqncom
\end{equation}
and similarly for $t\to b$. Correspondingly, $C^\dagger_{t/b,\pm}$ and $\aferm_{t/b,\pm}$ are related to $C$ and $\aferm$ via $U$ and $\tilde{U}$, respectively. 

\subsection{Explicit diagonalisation of the mass matrix\label{diagonalisation}}

We decompose the different fields with respect to their matrix elements in colour space as 
\begin{equation}
\label{eq: decomposition into blocks}
\begin{aligned}
 \Phi&=[\Phi]_{n,n'}E^{n}{}_{n'}+[\Phi]_{n,a}E^{n}{}_{a}+[\Phi]_{a,n}E^{a}{}_{n}+[\Phi]_{a,a'}E^{a}{}_{a'}
 \\&\phaneq+\Phi_{\text{tr}}((N-k)\idm_{k\times k}+k\idm_{(N-k)\times(N-k)})\eqncom
\end{aligned}
 \end{equation}
where $\Phi\in\{A_0,A_1,A_2,\scalq_4,\scalq_5,\scalq_6,c, C_{t,\pm}, C_{b,\pm}, \psi_{t,\pm}, \psi_{b,\pm}\}$, $n,n'=1,\dots,k$ and $a,a'=k+1,\dots,N$.
Moreover, we have split the diagonal components into individually traceless blocks, $\sum_n  [\Phi]_{n,n}=0=\sum_a  [\Phi]_{a,a}$, and a component $\Phi_{\text{tr}}$ proportional to the identity in each block.
Note that the matrix elements above are not independent degrees of freedom; apart from the aforementioned tracelessness condition, they are also (partially) related to each other via reality conditions.

The matrices $E^{a}{}_{a'}$ are annihilated by the $L_i$ and the corresponding components $[\Phi]_{a,a'}$ in the $(N-k)\times(N-k)$ block of all fields are hence massless. Moreover, the $L_i$ annihilate $((N-k)\idm_{k\times k}+k\idm_{(N-k)\times(N-k)})$ such that $\Phi_{\text{tr}}$ is also massless.

The matrices $E^{n}{}_{a}$ and $E^{a}{}_{n}$ in the off-diagonal $k\times(N-k)$ and $(N-k)\times k$ blocks transform in the irreducible $k$-dimensional representation of $\SU2$ with angular momentum $\ell=\frac{k-1}{2}$ and magnetic quantum number $m=\pm\left(\frac{k+1}{2}-n\right)$:
\begin{equation}
  L_i E^n{}_a = E^{n'}{}_a[t_i]_{n',n}\eqncom \qquad
  L_i E^a{}_n = -[t_i]_{n,n'}E^a{}_{n'}\eqndot
  \label{eq:Li-E}
\end{equation}
The same holds for the corresponding components of the fields.

The standard matrices $E^{n}{}_{n'}$ in the $k\times k$ block do not transform in an irreducible representation of $\SU2$ yet. The desired eigenstates yielding the decomposition to irreducible representations are provided by the spherical harmonics $\hat{Y}_\ell^m$ of the fuzzy sphere, where $\ell=1,\dots,k-1$ and $m=-\ell,\dots,\ell$.
They are explicitly given in appendix~\ref{fuzzy} and satisfy
\begin{equation}
  L_3 \hat{Y}_\ell^m = m \hat{Y}_\ell^m\eqncom\qquad
  L^2 \hat{Y}_\ell^m = \ell(\ell+1)\hat{Y}_\ell^m \eqndot
  \label{eq:fuzzy-eigenvalues}
\end{equation}
We thus write 
\begin{equation}
\label{eq: rewriting k x k}
 [\Phi]_{n,n'}E^{n}{}_{n'}=\Phi_{\ell,m}\hat{Y}_\ell^m\eqncom
\end{equation}
where the traceless $\hat{Y}_\ell^m$ implement the tracelessness condition $\sum_n[\Phi]_{n,n}=0$.
This concludes the diagonalisation of $L^2$.

For the easy bosons and ghosts, only $L^2$ occurs in the mass term, and $\Phi_{\ell,m}$, $[\Phi]_{n,a}$, $[\Phi]_{a,n}$, $[\Phi]_{a,a'}$ and $\Phi_{\text{tr}}$ completely diagonalise it. In terms of these components, the mass term reads
\begin{equation}
 -\frac{1}{2x_3^2}\tr(A_0L^2A_0)=-\frac{1}{2x_3^2}\left(2\frac{k^2-1}{4}[A_0]_{n,a}^\dagger [A_0]_{n,a}+\ell(\ell+1)(A_0)_{\ell,m}^\dagger (A_0)_{\ell,m}\right) \eqncom
\end{equation}
where we again have chosen $A_0$ for concreteness and used \eqref{eq:Ynorm used}. Here, $[A_0]_{n,a}^\dagger \equiv ([A_0]_{n,a})^\dagger= [A_0]_{a,n}$ and $(A_0)_{\ell,m}^\dagger\equiv((A_0)_{\ell,m})^\dagger=(-1)^m(A_0)_{\ell,-m}$.
Comparing this to the kinetic term
\begin{equation}
 -\frac{1}{2}\tr(A_0\partial^2A_0)=-\frac{1}{2}\left(2[A_0]_{n,a}^\dagger\partial^2[A_0]_{n,a}+(A_0)_{\ell,m}^\dagger\partial^2(A_0)_{\ell,m}\right)+\text{massless fields} \eqncom
\end{equation}
we immediately see that we have the nonzero mass eigenvalues $\frac{m^2}{x_3^2}=\frac{k^2-1}{4x_3^2}$ with multiplicity $2k(N-k)$ and $\frac{m^2}{x_3^2}=\frac{\ell(\ell+1)}{x_3^2}$ with multiplicity $2\ell+1$ for $\ell=1,\dots,k-1$. 
Note that in both equations we have used the first reality condition to remove $[A_0]_{a,n}$, resulting in the relative factor $2$ in front of the fields from the $k\times(N-k)$ block compared to those from the $k\times k$ block.

For the complicated bosons and the fermions, we have to diagonalise $J^2$ with $J_i=L_i+\frac{1}{2}\sigma_i$ in addition to $L^2$, see the discussion in the previous subsection.
Let $\Phi_{\pm}$ be a field with definite angular momentum $\ell$, magnetic quantum number $m$, spin $\frac{1}{2}$ and spin magnetic quantum number $\pm\frac{1}{2}$, i.e.\ $[C_{t,\pm}]_{n,a}$, $[C_{t,\pm}]_{a,n}$, $(C_{t,\pm})_{\ell,m}$ as well as the corresponding components of $C_{b,\pm}$, $\psi_{t,\pm}$, $\psi_{b,\pm}$, $\psi_{t,\pm}$ and $\psi_{b,\pm}$.  
The field can then be written in terms of the desired eigenstates of $L^2$ and $J^2$ as 
\begin{equation}
\begin{aligned}
  \Phi_{\pm}
    =&+ \left\langle j_1 = \ell,j_2 = \tfrac 1 2;m_1 = m, m_2 = \pm \tfrac 1 2\middle|
                     j = j_1 - \tfrac 1 2, m_j\right\rangle \Phi_{\aparallel,m_j}\\
     &+ \left\langle j_1 = \ell,j_2 = \tfrac 1 2;m_1 = m, m_2 = \pm \tfrac 1 2\middle|
                     j = j_1 + \tfrac 1 2, m_j \right\rangle \Phi_{\parallel,m_j}\eqndot
\end{aligned}                     
                \label{eq:heavy-light-CFields}     
\end{equation}
Here, $\Phi_{\aparallel,m_j}$ denotes the eigenstate with total angular momentum $j=\ell-\frac{1}{2}$ and $\Phi_{\parallel,m_j}$ denotes the eigenstate with total angular momentum $j=\ell+\frac{1}{2}$, i.e.\
\begin{equation}
\begin{aligned}
 L^2\Phi_{\aparallel,m_j}&=\ell(\ell+1)\Phi_{\aparallel,m_j}\eqncom &&
 L^2\Phi_{\parallel,m_j}=\ell(\ell+1)\Phi_{\parallel,m_j}\eqncom \\
 J^2\Phi_{\aparallel,m_j}&=(\ell-\tfrac{1}{2})(\ell+\tfrac{1}{2})\Phi_{\aparallel,m_j}\eqncom&&
 J^2\Phi_{\parallel,m_j}=(\ell+\tfrac{1}{2})(\ell+\tfrac{3}{2})\Phi_{\parallel,m_j}\eqndot
\end{aligned}
\end{equation}
The explicit expressions for the occurring Clebsch-Gordan coefficients are
\begin{align}
\label{eq: CG1}
  \left\langle j_1,j_2 = \tfrac 1 2;m_1, m_2 = + \tfrac 1 2\middle|
                     j = j_1 + \tfrac 1 2, m_j \right\rangle
  &= \delta_{m_j,m_1+m_2}\frac{\sqrt{j_1+m_1+1}}{\sqrt{2j_1+1}}\eqncom \\
  \left\langle j_1,j_2 = \tfrac 1 2;m_1, m_2 = - \tfrac 1 2\middle|
                     j = j_1 + \tfrac 1 2, m_j \right\rangle
  &= \delta_{m_j,m_1+m_2}\frac{\sqrt{j_1-m_1+1}}{\sqrt{2j_1+1}}\eqncom \\
  \left\langle j_1,j_2 = \tfrac 1 2;m_1, m_2 = + \tfrac 1 2\middle|
                     j = j_1 - \tfrac 1 2, m_j \right\rangle
  &= -\delta_{m_j,m_1+m_2}\frac{\sqrt{j_1-m_1}}{\sqrt{2j_1+1}}\eqncom
\intertext{and}
\label{eq: CG4}
\left\langle j_1,j_2 = \tfrac 1 2;m_1, m_2 = - \tfrac 1 2\middle|
                     j = j_1 - \tfrac 1 2, m_j \right\rangle
  &= \delta_{m_j,m_1+m_2}\frac{\sqrt{j_1+m_1}}{\sqrt{2j_1+1}}\eqndot
\end{align}

Using the above eigenstates, we can write the mass term of the complicated bosons as
\begin{multline}
 -\frac{1}{2x_3^2} \tr[C^T(L^2 - 4 S_i L_i) C]\\=-\frac{1}{2x_3^2}\biggl(
 2\frac{(k+2)^2-1}{4}C^\dagger_{at\aparallel,m_j}C_{at\aparallel,m_j} 
 +2\frac{(k-2)^2-1}{4}C^\dagger_{at\parallel,m_j}C_{at\parallel,m_j}  \\
 +(\ell^2+3\ell+2)C^\dagger_{\ell t\aparallel,m_j}C_{\ell t\aparallel,m_j} 
 +(\ell^2-\ell)C^\dagger_{\ell t\parallel,m_j}C_{\ell t\parallel,m_j}  
      + (t \to b)
 \biggr)     \eqncom
\end{multline}
where $C^\dagger_{at\aparallel,m_j}\equiv (C_{at\aparallel,m_j})^\dagger$, etc.
We have the (mostly) non-zero mass eigenvalues $\frac{m^2}{x_3^2}=\frac{(k+2)^2-1}{4x_3^2}$ with multiplicity $4(k-1)(N-k)$, $\frac{m^2}{x_3^2}=\frac{(k-2)^2-1}{4x_3^2}$ with multiplicity $4(k+1)(N-k)$, $\frac{m^2}{x_3^2}=\frac{\ell^2-\ell}{x_3^2}$ with multiplicity $4(\ell+1)$ and $\frac{m^2}{x_3^2}=\frac{\ell^2+3\ell+2}{x_3^2}$ with multiplicity $4\ell$ for $\ell=1,\dots,k-1$.

Similarly, we can write the fermion mass term as
\begin{equation}
 \begin{aligned}
  -\frac{1}{2x_3}\tr[\aferm G^iL_i\ferm]=-\frac{1}{2x_3}\biggl(&
 2\frac{k+1}{2}\aferm_{at\aparallel,m_j}\ferm_{at\aparallel,m_j} 
 -2\frac{k-1}{2}\aferm_{at\parallel,m_j}\ferm_{at\parallel,m_j}  \\
 &+(\ell+1)\aferm_{\ell t\aparallel,m_j}\ferm_{\ell t\aparallel,m_j} 
 -\ell\aferm_{\ell t\parallel,m_j}\ferm_{\ell t\parallel,m_j}  
      + (t \to b)
 \biggr)     \eqncom
 \end{aligned}
\end{equation}
where $\aferm_{at\aparallel,m_j}\equiv (\ferm_{at\aparallel,m_j})^\dagger\gamma^0$, etc.
In this case, we have the nonzero mass eigenvalues $\frac{m}{x_3}=\frac{k+1}{2x_3}$ with multiplicity $4(k-1)(N-k)$, $\frac{m}{x_3}=-\frac{k-1}{2x_3}$ with multiplicity $4(k+1)(N-k)$, $\frac{m}{x_3}=-\frac{\ell}{x_3}$ with multiplicity $4(\ell+1)$ and $\frac{m}{x_3}=\frac{\ell+1}{x_3}$ with multiplicity $4\ell$ for $\ell=1,\dots,k-1$.

\subsection{Summary of the spectrum}
\label{subsec: summary}

Defining 
\begin{equation}
\label{eq: definition nu}
 \nu=\sqrt{m^2+\frac{1}{4}}\eqncom
\end{equation}
we find the following pattern for the masses and $\nu$'s: \label{spectrum}
\begin{equation} \label{table:spectrum}
\begin{tabular}{c|c|c|c}
Multiplicity & $\nu(\scalq_{4,5,6},A_{0,1,2},c)$ & $m(\psi_{1,2,3,4})$ & $\nu(\scalq_{1,2,3},A_3)$ \\ \hline
$\ell +1$ & $\ell+\frac{1}{2}$ & $-\ell$ & $\ell-\frac{1}{2}$ \\
$\ell$ & $\ell+\frac{1}{2}$ & $\ell+1$ & $\ell+\frac{3}{2}$ \\
$(k+1)(N-k)$ & $\frac{k}{2}$ & $-\frac{k-1}{2}$ & $\frac{k-2}{2}$ \\
$(k-1)(N-k)$ & $\frac{k}{2}$ & $\frac{k+1}{2}$ & $\frac{k+2}{2}$ \\
$(N-k)(N-k)$ & $\frac{1}{2}$ & $0$ & $\frac{1}{2}$
\end{tabular}
\end{equation}
where $\ell=1,\ldots,k-1$. 

\section{Propagators \label{Propagators}}
Having diagonalised the quadratic part of the action, we can derive the propagators of the mass eigenstates.
Anticipating the use of dimensional regularisation and taking into account the symmetries of the
problem, we will work in $d+1$ dimensions with $d$ referring to the dimension of the codimension-one defect. 
For notational simplicity, we will
keep denoting the coordinate transverse to the defect as $x_3$.
We derive the scalar and fermionic propagators in subsections \ref{subsec: scalar propagators} and \ref{subsec: fermionic propagators}, respectively, by expressing them in terms of propagators in $AdS_{d+1}$.
We translate the propagators of the mass eigenstates to those of the flavour and colour eigenstates in subsection \ref{subsec: mixing part of propagators}.

\subsection{Scalar propagators}
\label{subsec: scalar propagators}
The scalar Minkowski space propagator $K(x,y)$ is the solution to
\begin{equation}
  \left(-\partial_\mu\partial^\mu+\frac{m^2}{x_3^2}\right)K(x,y) = \frac{g_\YM^2}{2}\delta(x-y)
  \eqncom
  \label{eq:prop-diff-eq}
\end{equation}
where the derivatives are all with respect to $x$, $\mu=0,1,\dots,d$ takes $d+1$ different values and $\frac{m}{x_3}$ is the ``mass'' coming from the classical expectation value.
The factor $g_\YM^2/2$ stems from the normalisation of the action in \eqref{eq:SYM-action}.

As noted in \cite{Nagasaki:2011ue}, $K(x,y)$ is basically the usual propagator
of a massive scalar in $AdS_{d+1}$. To see this, we write
\begin{equation}
  K(x,y) = \frac{g_\YM^2}{2}\frac{\tilde K(x,y)}{(x_3y_3)^{\frac{d-1}{2}}}\eqndot
\end{equation}
Equation \eqref{eq:prop-diff-eq} then becomes
\begin{equation}
\begin{aligned}
  \delta(x-y) &= \left(-\partial_\mu\partial^\mu+\frac{m^2}{x_3^2}\right)
  \frac{\tilde{K}(x,y)}{(x_3y_3)^{\frac{d-1}{2}}}\\
    &= \frac{1}{(x_3y_3)^{\frac{d-1}{2}}}\left(-\partial_\mu\partial^\mu
        +(d-1)\,\frac{1}{x_3}\partial_3 +\frac{m^2-\frac{d^2-1}{4}}{x_3^2}\right)\tilde{K}(x,y)
        \eqncom
\end{aligned}
\end{equation}
or
\begin{align}
  \left(-x_3^2\partial_\mu\partial^\mu+(d-1)x_3\partial_3+m^2-\frac{d^2-1}{4}\right)\tilde{K}(x,y) 
    = (x_3 y_3)^{\frac{d-1}{2}} x_3^2\,\delta(x-y) = x_3^{d+1}\delta(x-y) \eqndot
\end{align}
Let us now compare this to the $AdS_{d+1}$ case. We choose coordinates such that the (Euclidean) metric
is
\begin{equation}\label{AdSmetric}
  g_{\mu\nu} = \frac{1}{x_3^2}\delta_{\mu\nu} \eqncom \qquad
  g^{\mu\nu} = x_3^2\delta^{\mu\nu} \eqncom \qquad
  \sqrt{g} = \frac{1}{x_3^{d+1}} \eqndot
\end{equation}
The AdS propagator with mass $\tilde m$ is defined by
\begin{equation}
  (-\nabla_\mu\nabla^\mu + \tilde{m}^2)K_{\text{AdS}}(x,y)
    = \frac{\delta(x-y)}{\sqrt g}\eqndot
\end{equation}
Inserting the explicit expression \eqref{AdSmetric} for the metric, we find
\begin{equation}
\begin{aligned}
  x_3^{d+1}\delta(x-y) &= (-\nabla_\mu\nabla^\mu + \tilde{m}^2)K_{\text{AdS}}(x,y) \\
    &= -\frac{1}{\sqrt g}\partial_\mu(\sqrt{g} g^{\mu\nu} \partial_\nu K_{\text{AdS}}(x,y))
       +\tilde{m}^2 K(x,y)_{\text{AdS}}\\
    &= \bigl(-x_3^2\,\delta^{\mu\nu}\partial_\mu\partial_\nu
                +(d-1)x_3\partial_3+\tilde{m}^2\bigr)K_{\text{AdS}}(x,y) \eqndot
\end{aligned} 
\end{equation}
We see that the equations for $\tilde K(x,y)$ and $K_{\text{AdS}}(x,y)$ coincide, and hence
that
\begin{equation} \label{BosProp}
  K(x,y) = \frac{g_\YM^2}{2}\frac{\tilde K(x,y)}{(x_3y_3)^{\frac{d-1}{2}}} 
  = \frac{g_\YM^2}{2}\frac{K_{\text{AdS}}(x,y)}{(x_3y_3)^{\frac{d-1}{2}}}\eqncom
\end{equation}
with the identification 
\begin{equation}
\tilde m^2 = m^2 - \frac{d^2-1}{4}\eqndot 
\end{equation}
Notice that the above implies that the coordinate transverse to the defect, $x_3$, plays the role of
the radial coordinate of an $AdS_4$ space with the defect as its boundary. This interpretation continues to hold when fermions are taken into account, cf.~the next subsection. Notice also that none of the scalar masses in~\eqref{table:spectrum}
violate the Breitenlohner-Freedman (BF) bound~\cite{Breitenlohner:1982jf}, since $\tilde{m}^2 \geq -9/4$, which is precisely the BF bound for $AdS_4$. The  bound is saturated only for the special case $k= 2$.

Closed expressions  for
$K_{\text{AdS}}(x,y)$ in terms of hypergeometric functions can be found in
the literature, see e.g.~\cite{Allen:1985wd,Camporesi:1991nw}. Another representation, which is useful for our purpose, can be found
in~\cite{Liu:1998ty}, and reads
\begin{equation}\label{eq:PropagatorBessel}
\begin{aligned}
K_{\text{AdS}}(x,y)
&=(x_3 y_3)^{d/2} \int \frac{\de^d \vec{k}}{(2\pi)^d} \int _0^\infty \de w \frac{w}{w^2+\vec{k}^2}\e^{i \vec{k}\cdot(\vec{x}-\vec{y})} J_\nu (w x_3)J_\nu (w y_3),\\
&=(x_3 y_3)^{d/2} \int \frac{\de^d \vec{k}}{(2\pi)^d}\e^{i \vec{k}\cdot(\vec{x}-\vec{y})} I_{\nu}(|\vec{k}| x_3^<) K_{\nu}(|\vec{k}| x_3^>)\eqncom
\end{aligned} 
\end{equation}
where $I$ and $K$ are modified Bessel functions with $x_3^<$ ($x_3^>$) the smaller (larger) of $x_3$ and $y_3$ and $\nu$ was defined in \eqref{eq: definition nu}.

 \subsection{Fermionic propagators}   
 \label{subsec: fermionic propagators}
For the fermions, after diagonalisation and when working in Euclidean space where
$\{\gamma^\mu,\gamma^\nu\}=-2\delta^{\mu\nu}$, the propagator $K_F(x,y)$ fulfils 
\begin{equation}
  \left(-i\gamma^\mu\partial_\mu+\frac{m}{x_3}\right)K_F(x,y) = \frac{g_\YM^2}{2}\delta(x-y)
  \eqndot
\end{equation}
To relate this propagator to the propagator of fermions on $AdS_{d+1}$, we introduce
\begin{equation}
  K_F(x,y) = \frac{g_\YM^2}{2}\frac{\tilde K_F(x,y)}{(x_3)^{d/2}(y_3)^{d/2}}\eqndot
\end{equation}
Then, we find
\begin{equation}
\begin{aligned}
  \delta(x-y) &= \left(-i\gamma^\mu\partial_\mu+\frac{m}{x_3}\right)\frac{\tilde K_F(x,y)}{(x_3)^{d/2}(y_3)^{d/2}}\\
    &= \frac{1}{(y_3)^{d/2}}\left(-\frac{i}{(x_3)^{d/2}}\gamma^\mu\partial_\mu
        +\frac{d}{2}\,i\gamma^3\frac{1}{(x_3)^{d/2+1}} +\frac{m}{(x_3)^{d/2+1}}\right)\tilde{K}_F(x,y)
        \eqncom
\end{aligned}
\end{equation}
or
\begin{align}
  \left(-x_3i\gamma^\mu\partial_\mu+\frac{d}{2}i\gamma^3+m\right)\tilde{K}_F(x,y) 
    = (x_3)^{d/2+1} (y_3)^{d/2}\delta(x-y) = (x_3)^{d+1}\delta(x-y) \eqndot
\end{align}
Using again the AdS metric given in \eqref{AdSmetric}, the fermion propagator $K_{F,\text{AdS}}(x,y)$
solves 
\begin{equation}
  (-i\slash\hspace{-0.27cm}{D}  + \tilde{m})K_{F,\text{AdS}}(x,y)
    = \frac{\delta(x-y)}{\sqrt g}\eqncom
\end{equation}
where 
\begin{equation}
  \slash\hspace{-0.27cm}{D}= x_3 \partial_\mu \gamma^\mu -\frac{d}{2} \gamma^3
\end{equation}
is the spinor covariant derivative;
see \cite{Henningson:1998cd} and also \cite{Mueck:1998iz}.
Thus, we have
\begin{equation}
  K_F(x,y) = \frac{g_\YM^2}{2}\frac{\tilde K_F(x,y)}{(x_3)^{d/2}(y_3)^{d/2}}
   = \frac{g_\YM^2}{2}\frac{K_{F,\text{AdS}}(x,y)}{(x_3)^{d/2}(y_3)^{d/2}}\eqncom
\end{equation}
with $m=\tilde{m}$.

In \cite{Kawano:1999au}, the following useful expression for the fermionic propagator 
$K_{F,\text {AdS}}$ in $AdS_{d+1}$ in terms of the bosonic one is given:
\begin{align}
K_{F,\text {AdS}}^m(x,y) = \sqrt{\frac{y_3}{x_3}} \left [i\slash\hspace{-0.27cm}{D}+\frac{i}{2} \gamma^3+m\right]
\left[K_{\text {AdS}}^{\nu = m-\half}(x,y) {\cal P}_-+K_{\text {AdS}}^{\nu=m+\half}(x,y){\cal P}_+ \right ] \eqncom
\end{align}
where
\begin{equation}
{\cal P}_{\pm}=\frac{1}{2}(1\pm i\gamma^3) \eqndot
\end{equation}
From this, we can express the flat space fermionic propagator in terms of the bosonic one as follows
\begin{align}
K_F^m(x,y) &= x_3^{-\frac{d+1}{2}}\!\left [x_3i\gamma^\mu\partial_\mu-\frac{d-1}{2} i\gamma^3+m\right]
x_3^{\frac{d-1}{2}}\left[K^{\nu = m-\half}(x,y) {\cal P}_-+K^{\nu=m+\half}(x,y){\cal P}_+ \right ] 
\nonumber \\
&= \left [i\gamma^\mu\partial_\mu+\frac{m}{x_3}\right]\left[K^{\nu = m-\half}(x,y) {\cal P}_-+K^{\nu=m+\half}(x,y){\cal P}_+ \right ] \eqndot
\label{Finalfermionic}
\end{align}

For future reference, we note that the fermionic propagator enjoys the charge conjugation
symmetry
\begin{equation}
  \mc C \bigl(K_F(x,y)\bigr)^T \mc C^{-1} = K_F(y,x)\eqncom
  \label{eq:charge-conjugation-symmetry}
\end{equation}
where the transpose acts in spinor space, and $\mc C$ is defined 
in \eqref{eq:charge-conjugation-matrix}.

\subsection{Colour and flavour part of propagators}
\label{subsec: mixing part of propagators}

Using the mass eigenstates derived in section \ref{diagonalisation}, we can now rewrite the propagators of the fields with definite flavour in terms of the propagators of the mass eigenstates.

We begin with the fields in the $k \times k$ block. For the easy fields, the propagator is already
diagonal in the $\hat{Y}^m_\ell$ basis, so we have e.g.\
\begin{equation}
  \langle (\scalq_4)_{\ell,m}(x) (\scalq_4)_{\ell',m'}^\dagger(y) \rangle
    = \delta_{\ell,\ell'}\delta_{m,m'} K^{m^2 = \ell(\ell +1)}(x,y) \eqndot
\end{equation}
Here, $(\scalq_4)_{\ell,m}^\dagger\equiv((\scalq_4)_{\ell,m})^\dagger=(-1)^m(\scalq_4)_{\ell,-m}$ and $K^{m^2}$ is the propagator for a scalar mode with squared mass $m^2$, see section \ref{subsec: scalar propagators}.

Calculating the propagators for the complicated fields takes a little more effort.
It is useful to first consider the $C_{t,\pm}$ fields. Using the relation to the diagonal fields \eqref{eq:heavy-light-CFields} and suppressing space-time positions for brevity, we find
\begin{align}
  \langle (C_{t,+})_{\ell,m} (C_{t,+})_{\ell',m'}^\dagger \rangle
    &= \delta_{\ell,\ell'}\delta_{m,m'}
    \left(\frac{\ell+m+1}{2\ell+1}K^{m^2 = \ell(\ell-1)}+\frac{\ell-m}{2\ell+1}K^{m^2 = (\ell+1)(\ell+2)}\right)\eqncom\\
  \langle (C_{t,-})_{\ell,m} (C_{t,-})_{\ell',m'}^\dagger \rangle
    &= \delta_{\ell,\ell'}\delta_{m,m'}
    \left(\frac{\ell-m+1}{2\ell+1}K^{m^2 = \ell(\ell-1)}+\frac{\ell+m}{2\ell+1}K^{m^2 = (\ell+1)(\ell+2)}\right)\eqncom\\
  \langle (C_{t,+})_{\ell,m} (C_{t,-})_{\ell',m'}^\dagger \rangle
    &= \delta_{\ell,\ell'}\frac{[t_-^{(2\ell+1)}]_{\ell-m+1,\ell-m'+1}}{2\ell+1}(
    K^{m^2 = \ell(\ell-1)}-K^{m^2 = (\ell+1)(\ell+2)})\eqncom
\intertext{and}
  \langle (C_{t,-})_{\ell,m} (C_{t,+})_{\ell',m'}^\dagger \rangle
    &= \delta_{\ell,\ell'}\frac{[t_+^{(2\ell+1)}]_{\ell-m+1,\ell-m'+1}}{2\ell+1}(
    K^{m^2 = \ell(\ell-1)}-K^{m^2 = (\ell+1)(\ell+2)})\eqndot
\end{align}
Here, $t^{(2\ell+1)}_i$ are the generators of the $(2\ell+1)$-dimensional irreducible representation of the Lie algebra
$\SU{2}$ defined in appendix~\ref{representationmatrices} with $k \to 2\ell+1$.
The propagators with $t\to b$ are identical, while the mixed ones vanish.
Using \eqref{eq: def top and bottom fields}, we express the original fields
in terms of $C_{t,\pm}$ and $C_{b,\pm}$. We can now compute e.g.\
\begin{equation}
 \begin{aligned}
  \langle (\scalq_1)_{\ell,m} (\scalq_2)_{\ell',m'}^\dagger \rangle
    &= \frac{1}{2}\left(-i\langle (C_{t,+})_{\ell,m} (C_{t,+})_{\ell',m'}^\dagger \rangle + i\langle (C_{b,-})_{\ell,m} (C_{b,-})_{\ell',m'}^\dagger \rangle \right)\\
    &= -i\delta_{\ell,\ell'}
    \frac{[t_3^{(2\ell+1)}]_{\ell-m+1,\ell-m'+1}}{2\ell+1}(K^{m^2 = \ell(\ell-1)}-K^{m^2 = (\ell+1)(\ell+2)})\eqndot
\end{aligned}
\end{equation}
Repeating this exercise, we finally find
\begin{align}
\label{eq:HardScalarLMpropagator}
  \langle (\scalq_i)_{\ell,m} (\scalq_j)_{\ell',m'}^\dagger \rangle
    &= \delta_{i,j}\delta_{\ell,\ell'}\delta_{m,m'}
    \left(\frac{\ell+1}{2\ell+1}K^{m^2 = \ell(\ell-1)}+\frac{\ell}{2\ell+1} K^{m^2 = (\ell+1)(\ell+2)}\right)
    \\
    &\phaneq   -i\teps_{ijl}[t_l^{(2\ell+1)}]_{\ell-m+1,\ell-m'+1}\delta_{\ell,\ell'}\frac{1}{2\ell+1}
    (K^{m^2 = \ell(\ell-1)}-K^{m^2 = (\ell+1)(\ell+2)})\eqncom\nonumber\\
  \langle (A_3)_{\ell,m} (A_3)_{\ell',m'}^\dagger \rangle
    &= \delta_{\ell,\ell'}\delta_{m,m'}
    \left(\frac{\ell+1}{2\ell+1}K^{m^2 = \ell(\ell-1)}+\frac{\ell}{2\ell+1} K^{m^2 = (\ell+1)(\ell+2)}\right)
\intertext{and}
  \langle (\scalq_i)_{\ell,m} (A_3)_{\ell',m'}^\dagger \rangle
  &= -\langle (A_3)_{\ell,m} (\scalq_i)_{\ell',m'}^\dagger \rangle\\
  &= i \delta_{\ell,\ell'}\frac{[t_i^{(2\ell+1)}]_{\ell-m+1,\ell-m'+1}}{2\ell+1}
  (K^{m^2 = \ell(\ell-1)}-K^{m^2 = (\ell+1)(\ell+2)})\eqndot \nonumber
\end{align}

Similarly, we obtain the propagators of the fermions as
\begin{equation}\label{eq:FermionDiagProp}
\begin{aligned}
  \langle (\ferm_i)_{\ell,m} \overline{\vphantom{\aferm}(\ferm_j)}_{\ell',m'} \rangle
    &= \delta_{i,j}\delta_{m,m'}\delta_{\ell,\ell'}
        \left(\frac{\ell+1}{2\ell+1}K_F^{m=-\ell}+\frac{\ell}{2\ell+1}K_F^{m=\ell+1}\right)\\
    &\phaneq -\delta_{\ell,\ell'} [G^l]_{i,j}
        \frac{[t_l^{(2\ell+1)}]_{\ell-m+1,\ell-m'+1}}{2\ell+1}
        \left(K_F^{m=-\ell}-K_F^{m=\ell+1}\right)
        \eqncom
\end{aligned}
\end{equation}
where $\overline{\vphantom{\aferm}(\ferm_j)}_{\ell',m'}\equiv((\ferm_j)_{\ell',m'})^\dagger \gamma^0=(-1)^{m'}(\aferm_j)_{\ell',-m'}$, $G^l$ are the $4\times 4$ 
matrices defined in (\ref{eq:G-matrices})
and $K_F^m$ denotes the fermionic propagators of definite mass $m$ derived in section \ref{subsec: fermionic propagators}.

To obtain the propagator between the matrix elements, one can write
\begin{equation}
  \langle [\Phi_1]_{n_1,n_2} [\Phi_2]_{n_3,n_4} \rangle
    = [\hat{Y}^m_\ell]_{n_1,n_2}[(\hat{Y}^{m'}_{\ell'})^\dagger]_{n_3,n_4} 
       \langle (\Phi_1)_{\ell,m} (\Phi_2)_{\ell',m'}^\dagger \rangle
\end{equation}
and use \eqref{eq: matrix element Y} to get an explicit expression. In practice, however, it is often more
convenient to work directly in the $\hat{Y}_\ell^m$ basis.

We have now written all the propagators for the $k\times k$ block. To obtain the corresponding expressions
for the $k\times (N-k)$ and $(N-k) \times k$ blocks is mostly a matter of replacing
$(\Phi)_{\ell,m} \to [\Phi]_{n,a}$ and $\ell \to (k-1)/2$ in the above formulae. In particular, we have
\begin{align}
\label{eq: propagator off-diagonal easy scalars}
  \langle [\scalq_4]_{n,a} [\scalq_4]_{n',a'}^\dagger \rangle
    &= \delta_{n,n'}\delta_{a,a'} K^{m^2 = \frac{k^2-1}{4}} \eqncom\\
   \langle [A_3]_{n,a} [A_3]_{n',a'}^\dagger \rangle
    &= \delta_{n,n'}\delta_{a,a'}\left(\frac{k+1}{2k} K^{m^2 =\frac{(k-2)^2-1}{4}}+\frac{k-1}{2k} K^{m^2 = \frac{(k+2)^2-1}{4}}\right)\eqncom\\
\label{eq:PropHardOffDiag}
  \langle [\scalq_i]_{n,a} [\scalq_j]_{n',a'}^\dagger \rangle
    &= \delta_{i,j}\delta_{n,n'}\delta_{a,a'}\left(\frac{k+1}{2k} K^{m^2 =\frac{(k-2)^2-1}{4}}+\frac{k-1}{2k} K^{m^2 = \frac{(k+2)^2-1}{4}}\right)\\\nonumber
    &\phaneq   -i\teps_{ijl}[t_l]_{n,n'}\delta_{a,a'}\frac{1}{k}\left(K^{m^2 =\frac{(k-2)^2-1}{4}}-K^{m^2 = \frac{(k+2)^2-1}{4}}\right)\eqncom\\
  \langle [\scalq_i]_{n,a} [A_3]_{n',a'}^\dagger \rangle &= - \langle [A_3]_{n,a} [\scalq_i]_{n',a'}^\dagger \rangle
  = i  [t_i]_{n,n'}\delta_{a,a'}\frac{1}{k}\left(K^{m^2 =\frac{(k-2)^2-1}{4}}-K^{m^2 = \frac{(k+2)^2-1}{4}}\right)
\intertext{and}
\label{eq:FermionOffDiagProp}
 \langle [\ferm_i]_{n,a} \overline{\vphantom{\aferm}[\ferm_j]}_{n',a'} \rangle
    &= \delta_{a,a'}\delta_{i,j}\delta_{n,n'}\frac{1}{k}
	        \left(\frac{k+1}{2}K_F^{m=-\frac{k-1}{2}}
	                +\frac{k-1}{2}K_F^{m=\frac{k+1}{2}}\right)\\\nonumber
    &\phaneq -\delta_{a,a'} [G^l]_{i,j}
        \frac{[t_l]_{n,n'}}{k}
        \left(K_F^{m=-\frac{k-1}{2}}-K_F^{m=\frac{k+1}{2}}\right)\eqncom
\end{align}
where $[\scalq_4]_{n',a'}^\dagger \equiv([\scalq_4]_{n',a'})^\dagger =[\scalq_4]_{a',n'}$, $\overline{\vphantom{\aferm}[\ferm_j]}_{n',a'}\equiv([\ferm_j]_{n',a'})^\dagger \gamma^0=[\aferm_j]_{a',n'}$, etc.

Fermionic propagators with bars added and/or removed can be obtained from those
given above using the 
Majorana condition $\ferm_i = \mc C \aferm_i^T$; see appendix~\ref{app: 10Dto4D}. In particular, we will need the propagator
\begin{align}
 \langle [\ferm_i]_{a,n} \overline{\vphantom{\aferm}[\ferm_j]}_{a',n'} \rangle
    &= \delta_{a,a'}\delta_{i,j}\delta_{n,n'}\frac{1}{k}
	        \left(\frac{k+1}{2}K_F^{m=-\frac{k-1}{2}}
	                +\frac{k-1}{2}K_F^{m=\frac{k+1}{2}}\right)\\\nonumber
    &\phaneq +\delta_{a,a'} [G^l]_{i,j}
        \frac{[t_l]_{n',n}}{k}
        \left(K_F^{m=-\frac{k-1}{2}}-K_F^{m=\frac{k+1}{2}}\right)\eqndot
\end{align}
Here, we have used the charge conjugation symmetry \eqref{eq:charge-conjugation-symmetry}
to simplify the expression.

\section{Dimensional regularisation\label{dimreg}}

 For our one-loop computation, we need to evaluate $K(x,x)$ as well as $\tr K_F(x,x)$ and we hence need to regulate these quantities. Dimensional  regularisation  has been used successfully in combination with dimensional 
 reduction in a number of higher loop computations in standard ${\cal N}=4$ SYM theory, see e.g.~\cite{Erickson:2000af,Nandan:2014oga} and references therein,
but neither have been tested in the defect setup.
 In this section, we determine $K(x,x)$ as well as $\tr K_F(x,x)$ in dimensional regularisation and discuss the preservation of supersymmetry in analogy to dimensional reduction. 

Results for  $K(x,x)$ and  $\tr K_F(x,x)$ in Hadamard as well as zeta-function regularisation,
which are commonly used in AdS, can be found in the literature and for completeness we summarise these in appendix~\ref{Hadamard}.

\paragraph{Bosonic fields}
In order to evaluate $K(x,x)$ using dimensional regularisation, we use as our starting point the 
expression~(\ref{eq:PropagatorBessel}), consider the 
$\vec{k}$ integral in $d=3-2\peps$ dimensions, set $\vec{x}=\vec{y}$ and go to polar coordinates. The expression \eqref{BosProp} then
turns into
\begin{equation}
\begin{aligned}
K^{m^2=\nu^2-\frac{1}{4}}(x,x)
&=
\frac{g_\YM^2}{2}\, \,x_3\,\, \frac{2 \pi^{3/2-\peps}}{\Gamma(3/2-\peps)}
 \int_0^{\infty} \de k\, \frac{k^{2-2\peps} }{(2\pi)^{3-2\peps}} I_{\nu}({k} x_3) K_{\nu}(k x_3)\eqncom
\end{aligned} 
\end{equation}
where $k$ denotes the radial component of $\vec{k}$ and $\frac{2 \pi^{3/2-\peps}}{\Gamma(3/2-\peps)}$ is the area of the unit sphere
in $d=3-2\peps$ dimensions resulting from the angular integration. Expanding in small $\peps$ and dropping terms of $O(\peps)$, we find 
\begin{equation}
\int_0^{\infty} \de k\, {k^{2-2\peps} } I_{\nu}({k} x_3) K_{\nu}(k x_3)=
 \frac{1}{8\, x_3^3} \left(2 m^2\left[\frac{1}{2}+ \digamma(\nu+\half)  -\log 2x_3 - \frac{1}{2\peps}\right]-1\right) \eqndot 
\end{equation}
This means that the total, regularised propagator is given by 
\begin{align}\label{eq:DimRegBoson}
K^{\nu}(x,x) =\frac{g_\YM^2}{2}\,\,
 \frac{1}{16\pi^2\, x_3^2} \left( m^2 \left[- \frac{1}{\peps} -\log(4\pi)+\gammaE  -2\log (x_3)+ 2\digamma(\nu+\half)-1  \right]-1\right)\eqncom
\end{align}
where $\gammaE$ is the Euler-Mascheroni constant.

The form of the bosonic spectrum found in the previous section means that the digamma function $\digamma$ simplifies. We first observe that the eigenvalues come in two
families. The first family is 
\begin{equation}
  m^2 = \frac{(k+2s)^2-1}{4}\eqncom\qquad s\in\{-1,0,1\}\eqncom
\end{equation}
and the second family is
\begin{equation}
  m^2 = j(j-1),\qquad j=1,\ldots,k+1\eqndot
\end{equation}
The digamma terms then reduce to
\begin{align}
  \digamma\left(\sqrt{\frac{(k+2s)^2-1}{4}+\frac{1}{4}}+\frac{1}{2}\right)
    &= \begin{cases}
   -\gammaE-2\log 2
       +\sum_{n=1}^{\frac{k}{2}+s}\frac{2}{2n-1}\eqncom & k\text{ even}\eqncom \\
       -\gammaE
       +\sum_{n=1}^{\frac{k-1}{2}+s}\frac{1}{n}\eqncom & k\text{ odd}\eqncom
       \end{cases} \\
\intertext{and}
  \digamma\left(\sqrt{j(j-1)+\frac{1}{4}}+\frac{1}{2}\right)
    &= -\gammaE+\sum_{n=1}^{j-1}\frac{1}{n}\eqncom
\end{align}
respectively.

\paragraph{Fermionic fields}
The other quantity that is relevant for our one-loop computation is the trace
of the fermionic propagator.  In this case, we will use as our starting point the formula~(\ref{Finalfermionic}).
Since the $\gamma$ matrices are traceless and furthermore satisfy $\tr (\gamma^i\gamma^3)=0$,  
what remains to evaluate is then effectively
\begin{align}
\tr K_F^m(x,y) &= 2\left [-\partial_3 + \frac{m}{x_3}\right]K^{\nu = m-\half}(x,y) 
 +2\left [\partial_3 + \frac{m}{x_3}\right]K^{\nu=m+\half}(x,y) \eqncom
\end{align}
where we have used that $\tr{m}=4 \,m$ and $\tr (\gamma^3)^2=-4$. Now, we have to find the regularised version
of this expression at coinciding points, $K_F(x,x)$. 

Using the fact that $\tr K_F(x,y)$ and $K(x,y)$ are symmetric under interchanging $x$ and $y$,%
\footnote{For $\tr K_F(x,y)$, this follows from \eqref{eq:charge-conjugation-symmetry}.}
we can write 
\begin{equation}
 \begin{aligned}
\tr K_F^m(x,y) &= \left [-\partial_{x_3}-\partial_{y_3} + \frac{m}{x_3}+ \frac{m}{y_3}\right]K^{\nu = m-\half}(x,y) \\
&\phaneq+ \left [\partial_{x_3}+\partial_{y_3} + \frac{m}{x_3}+ \frac{m}{y_3}\right]K^{\nu=m+\half}(x,y) \eqndot
\end{aligned}
\end{equation}
In the limit $y\to x$, we have $(\partial_{x_3}+\partial_{y_3})K(x,y)\to \partial_{x_3}K(x,x)$, such that 
\begin{align}
\tr K_F^m(x,x) &= \left [-\partial_{x_3}+ 2\frac{m}{x_3}\right]K^{\nu = m-\half}(x,x) 
+ \left [\partial_{x_3} +  2\frac{m}{x_3}\right]K^{\nu=m+\half}(x,x) \eqndot
\end{align}
Substituting the regularised expression \eqref{eq:DimRegBoson} for the boson into this then leads to
\begin{multline}
\tr K_F^m(x,x) =\frac{g_\YM^2}{2}\, \frac{1}{4\pi^2x_3^3}\Bigg[ m^3+m^2-3m-1\label{Fermionsdimreg}\\
+m(m^2-1)\left(-\frac{1}{\peps} -\log(4\pi)+\gammaE- 2\log (x_3) +2 \digamma(m)-2\right)\Bigg]\eqndot 
\end{multline}

The diagonalisation of the fermionic mass terms yields both positive and negative eigenvalues.
By chirally rotating the fermion fields, one can argue that the sign of the mass should 
only affect the overall sign of the fermion loop; cf.\ also the expression for the 
propagator in \cite{Muck:1999mh}.
Hence, the full $m$ dependence of~\eqref{Fermionsdimreg} is
\begin{multline}
   \tr K_F^m(x,x) =\sgn(m)\, \frac{g_\YM^2}{2}\,\frac{1}{4\pi^2x_3^3}\Bigg[ |m|^3+|m|^2-3|m|-1  \\ 
+|m|(|m|^2-1)\left(-\frac{1}{\peps} -\log(4\pi)+\gammaE-2 \log (x_3) +2\digamma(|m|)-2\right)\Bigg] \eqndot
 \end{multline}

\paragraph{Dimensional reduction}

Dimensional regularisation alone breaks supersymmetry, as the number of components of the gauge field $A^\mu$ is changed from $n_A=4$ to $n_A=D=4-2\peps$ while the numbers of fermions $n_\psi=4$ and real scalars $n_\phi=6$ remains unchanged.
In usual $\mathcal{N}=4$ SYM theory, a supersymmetry-preserving alternative to dimensional regularisation is dimensional reduction \cite{Siegel:1979wq,Capper:1979ns}.%
\footnote{Note that dimensional reduction is inconsistent at sufficiently high loop orders though \cite{Siegel:1980qs,Avdeev:1981vf,Avdeev:1982np,Avdeev:1982xy}.}
It uses the fact that $\mathcal{N}=4$ SYM theory in four dimensions is the dimensional reduction of $\mathcal{N}=1$ SYM theory in ten dimensions. Dimensionally reducing to $D=4-2\peps$ dimensions instead leads to a supersymmetry-preserving regularisation with $n_\psi=4$ fermions but $n_\phi=6+2\peps$ real scalars.

Our regularisation will follow the spirit of dimensional reduction adapted to the situation with the defect and the classical 
vevs.
In our dCFT, the gauge fields and scalars are split into easy and complicated fields: $n_A=n_{A,\text{easy}}+n_{A,\complicated}=4-2\peps$ and $n_\phi=n_{\phi,\text{easy}}+n_{\phi,\complicated}=6+2\peps$. In the calculation above, we have only touched the $d$ dimensions parallel to the defect, such that the codimension of the defect remains one. Thus, we have $n_{A,\text{easy}}=3-2\peps$ and $n_{A,\complicated}=1$. Furthermore, 
we have left untouched the three scalar fields which acquire vevs as this ensures that the classical equations
of motion and the Nahm condition which define the fuzzy-funnel solution continue to be fulfilled away from $d=3$.
Thus, we are led to conclude $n_{\phi,\complicated}=3$ and $n_{\phi,\text{easy}}=3+2\peps$.

Further support for the above conclusion comes from the construction via the D5-D3 probe-brane set-up.
The easy gauge fields corresponds to the directions in which both the D5 and the D3 brane extend, while the easy scalars correspond to the directions into which none of the branes extend. The complicated scalars (gauge field) correspond to the directions in which only the D5 (D3) extends.
For the D5-D3 probe-brane set-up, supersymmetry requires that the number of Neumann-Dirichlet directions, i.e.\ the number of dimensions in which only the D5 brane \emph{or} the D3 branes extend, is 0, 4 or 8; see for instance \cite{Polchinski:1998rr,Ammon:2015wua}.
Thus, supersymmetry requires that we further keep $n_{A,\complicated}+n_{\phi,\complicated}=10-n_{A,\text{easy}}+n_{\phi,\text{easy}}=4$ fixed, which indeed leads to $n_{\phi,\complicated}=3$ and $n_{\phi,\text{easy}}=3+2\peps$.

\section{One-loop corrections to one-point functions\label{oneloop}}

For operators $\mathcal{O}$ with definite scaling dimension $\Delta$, conformal symmetry constrains the one-point function to be of the form \cite{Cardy:1984bb} 
\begin{equation}
\langle {\cal O}_{\Delta}\rangle(x)=\frac{C}{x_3^{\Delta}}\eqncom
\end{equation}
where $C$ is a constant and $x_3$ denotes the distance to the defect.

\begin{figure}[t]
\centering
 \subfigure[]{
\centering
  \includegraphics{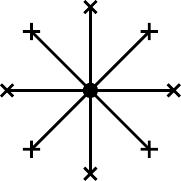}%
  \label{subfig: tree}
} \qquad 
 \subfigure[]{
\centering
  \includegraphics{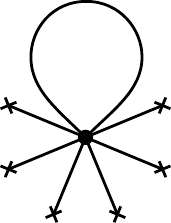}%
  \label{subfig: tadpole}
} \qquad 
 \subfigure[]{
\centering
  \includegraphics{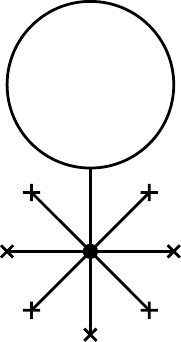}%
  \label{subfig: lolipop}
}
\caption{The diagrams which contribute to the one-point functions of scalar fields at tree level~\subref{subfig: tree} and one-loop order (\subref{subfig: tadpole} tadpole and \subref{subfig: lolipop} lollipop).  The operator is represented by a dot and
a cross symbolises the insertion of the classical solution.\label{fig: one-point functions}}
\end{figure}

Let us consider a general single-trace operator built out of $L$ real scalars:
\begin{equation}
\label{eq: general operator}
 \mathcal{O}(x)=\mathcal{O}^{i_1i_2\dots i_L}\tr(\scal_{i_1}\scal_{i_2}\dots \scal_{i_L})(x)\eqndot
\end{equation}
The classical one-point function is simply given by inserting the classical solution \eqref{phicl} into \eqref{eq: general operator}:
\begin{equation}
\label{eq: one-point general operator tree level}
 \langle\mathcal{O}\rangle_{\text{tree}}(x)=\mathcal{O}^{i_1i_2\dots i_L}\tr(\scalc_{i_1}\scalc_{i_2}\dots \scalc_{i_L})(x)\eqndot
\end{equation}
This is depicted in figure \ref{subfig: tree}.
We now calculate the first quantum correction to this quantity.

\subsection{One-loop one-point functions of general operators}

At one-loop order, two different diagrams can contribute to the one-point function of any operator. We call them the lollipop diagram and the tadpole diagram and depict them in figure \ref{subfig: lolipop} and \ref{subfig: tadpole}, respectively.

The lollipop diagram is obtained by expanding the operator to linear order in the quantum fields and connecting this quantum field with a propagator to a quantum field in a cubic vertex whose other two quantum fields are connected with each other by a second propagator:
\begin{equation}
\label{eq: general lollipop diagram}
 \langle\mathcal{O}\rangle_{\text{1-loop,lol}}(x)=\mathcal{O}^{i_1i_2\dots i_L}
 \sum_{j=1}^L\tr(\scalc_{i_1}\dots \contraction{}{\scalq}{_{i_j}\dots \scalc_{i_L})(x)\int\de^4y\sum_{\Phi_1,\Phi_2,\Phi_3}V_3(}{\Phi}\scalq_{i_j}\dots \scalc_{i_L})(x)\int\de^4y\sum_{\Phi_1,\Phi_2,\Phi_3}V_3(\Phi_1,\contraction{}{\Phi}{_2,}{\Phi}\Phi_2,\Phi_3)(y)\eqncom
\end{equation}
where the second sum is over all cubic vertices $V_3$ in the theory.
Note that this diagram is 1-particle-reducible and effectively is expressed in terms of the contribution of the one-loop correction to the scalar vevs:
\begin{align}
\label{eq: general lollipop diagram reformulated}
 \langle\mathcal{O}\rangle_{\text{1-loop,lol}}(x)&=\mathcal{O}^{i_1i_2\dots i_L}
 \sum_{j=1}^L\tr(\scalc_{i_1}\dots\langle \scal_{i_j}\rangle_{\text{1-loop}}\dots \scalc_{i_L})(x) \eqncom
 \intertext{where}
 \label{eq: general lollipop diagram reformulated 2}
 \langle \scal_i\rangle_{\text{1-loop}}(x)&=\contraction{}{\scalq}{_{i}(x)\int\de^4y\sum_{\Phi_1,\Phi_2,\Phi_3}V_3(}{\Phi}\scalq_{i}(x)\int\de^4y\sum_{\Phi_1,\Phi_2,\Phi_3}V_3(\Phi_1,\contraction{}{\Phi}{_2,}{\Phi}\Phi_2,\Phi_3)(y)
 \eqndot
\end{align}
We calculate $\langle \scal_i\rangle_{\text{1-loop}}$ in appendix \ref{app: one-loop vevs}, finding
\begin{equation}
\label{eq: correction to scalar vev main text}
 \langle \scal_i\rangle_{\text{1-loop}}(x)= 0
 \eqndot
\end{equation}
Thus,
\begin{align}
\label{eq: general lollipop diagram final}
 \langle\mathcal{O}\rangle_{\text{1-loop,lol}}(x)=0
 \eqncom
\end{align}
independently of which operator we are looking at.

The tadpole diagram is obtained by expanding the operator to quadratic order in the quantum fields and connecting the resulting two quantum fields with a propagator:
\begin{equation}
 \langle\mathcal{O}\rangle_{\text{1-loop,tad}}(x)=\sum_{j_1,j_2}\mathcal{O}^{i_1\dots i_{j_1} \dots i_{j_2} \dots i_L}\tr(\scalc_{i_1}\dots \contraction{}{\scalq}{_{i_{j_1}}\dots }{\scalq}\scalq_{i_{j_1}}\dots \scalq_{i_{j_2}}\dots\scalc_{i_L})(x)\eqndot
\end{equation}
In the large-$N$ limit, the tadpole integral only contributes when the two quantum fields are neighbouring, i.e.\ when $j\equiv j_1=j_2-1$; the components in the off-diagonal $k\times (N-k)$ and $(N-k)\times k$ blocks can contribute only in this case, 
and only they scale with $N$.\footnote{Recall that the fields in the $(N-k)\times(N-k)$ block do not directly couple to the classical fields. Moreover, they are massless such that their tadpole integrals vanish.}
Inserting the decomposition \eqref{eq: decomposition into blocks}, we find 
\begin{equation}
\label{eq: general tadpole result}
\begin{aligned}
 \langle\mathcal{O}\rangle_{\text{1-loop,tad}}(x)=&\sum_{j}\mathcal{O}^{i_1\dots i_j\, i_{j+1} \dots i_L}\tr(\scalc_{i_1}\dots E^n{}_aE^a{}_{n'}\dots\scalc_{i_L})(x)\langle[\scalq_{i_j}]_{n,a}(x)[\scalq_{i_{j+1}}]_{a,n'}(x)\rangle\\
 &+(k\times k)\text{-contributions} \eqndot
 \end{aligned}
\end{equation}
The occurring propagator is only non-vanishing for $i_j=i_{j+1}=4,5,6$ and $i_j,i_{j+1}=1,2,3$.
All required cases are given in subsection \ref{subsec: mixing part of propagators}.

At one-loop order, the one-point functions do not receive contributions from the quartic vertices as the occurrence of such a
vertex would require an additional propagator in comparison with a cubic vertex. The one-point functions do not receive any contributions from the fields living on the defect either. This is due to the fact that any such one-loop diagram would involve a loop consisting
of a single propagator of a defect field, which vanishes due to conformal invariance.  

In general, there are two further contributions at one-loop level. 
The first originates from the need to renormalise the operator via the renormalisation constant $\mathcal{Z}=1+\mathcal{Z}_{\text{1-loop}}+O(\lambda^2)$:
\begin{equation}
\label{eq: general result renormalisation}
\begin{aligned}
 \langle\mathcal{O}\rangle_{\text{1-loop},\mathcal{Z}}(x)=&\langle\mathcal{Z}_{\text{1-loop}}\mathcal{O}\rangle_{\text{tree}}(x) \eqndot
 \end{aligned}
\end{equation}
This contribution cancels the UV divergence in \eqref{eq: general tadpole result}, see also the discussion underneath \eqref{eq: one-loop renormalisation constant contribution SU(2)}. 
The second additional contribution originates from the first quantum correction to the one-loop eigenstate, i.e.\ the two-loop eigenstate, if we are looking at operators of definite scaling dimension $\Delta$:
\begin{equation}
\label{eq: one-point general operator eigenstate}
 \langle\mathcal{O}\rangle_{\text{1-loop},\mathcal{O}}(x)=\mathcal{O}_{\text{2-loop}}^{i_1i_2\dots i_L}\tr(\scalc_{i_1}\scalc_{i_2}\dots \scalc_{i_L})(x)\eqndot
\end{equation}

Thus, we have for the planar one-loop one-point function of \emph{any} single-trace operator built out of scalar fields:
\begin{equation}
\label{eq: one-point general operator total}
 \langle\mathcal{O}\rangle_{\text{1-loop}}(x)=
 \langle\mathcal{O}\rangle_{\text{1-loop},\text{tad}}(x)
 +\langle\mathcal{O}\rangle_{\text{1-loop},\mathcal{Z}}(x)
 +\langle\mathcal{O}\rangle_{\text{1-loop},\mathcal{O}}(x)
 \eqndot
\end{equation}

\subsection{One-loop one-point functions in the \texorpdfstring{$\SU2$}{SU(2)} sector\label{sec:SU(2)}}

Let us now consider operators in the $\SU2$ sector, which are built from the complex scalars $\Phi_\downarrow\equiv X=\scal_1+i\scal_4$ and $\Phi_\uparrow\equiv Z=\scal_{3}+i\scal_6$.
Consider the operator
\begin{equation}
 \mathcal{O}(x)=\mathcal{O}^{s_1s_2\dots s_L}\tr(\Phi_{s_1}\Phi_{s_2}\dots \Phi_{s_L})(x)\eqncom
\end{equation}
where $s_i=\uparrow,\downarrow$.
The tree-level one-point functions of these operators were computed using integrability in \cite{deLeeuw:2015hxa, Buhl-Mortensen:2015gfd}.

Of the above diagrams contributing to the one-loop one-point function, only the tadpole diagram simplifies further if we 
restrict ourselves to the $\SU2$ sector. Using the explicit expressions for the propagators given in section \ref{subsec: mixing part of propagators}, we find
\begin{align}
  \langle\mathcal{O}\rangle_{\text{1-loop,tad}}(x)&=\frac{\lambda}{16\pi^2}\frac{1}{(x_3)^2}\sum_{j}\delta_{s_j=s_{j+1}}\mathcal{O}^{s_1\dots s_j\, s_{j+1} \dots s_L}\tr(\scalc_{s_1}\dots\scalc_{s_{j-1}} \scalc_{s_{j+2}}\dots\scalc_{s_L})(x)\nonumber\\
&\phaneq +\frac{\lambda}{8\pi^2} 
\left(
 -\frac{1}{2 \peps} - \frac{1}{2}\log(4\pi)+\frac{1}{2}\gammaE - \log (x_3)+\digamma(\tfrac{k+1}{2}) \right)\\
 &\phaneq\phaneq\times
\sum_{j}\mathcal{O}^{s_1\dots s_j\, s_{j+1} \dots s_L}\tr(\scalc_{s_1}\dots\scalc_{s_{j-1}} [\scalc_{s_{j}},\scalc_{s_{j+1}}]\scalc_{s_{j+2}}\dots\scalc_{s_L})(x)\eqndot\nonumber
 \end{align}
We observe that the third line is precisely proportional to the one-loop dilatation operator in the $\SU{2}$ sector originally obtained in \cite{Minahan:2002ve}.
For one-loop eigenstates, the third line is proportional to the one-loop anomalous dimension multiplied by the tree-level one-point function:
\begin{align}
  &\langle\mathcal{O}\rangle_{\text{1-loop,tad}}(x)=\frac{\lambda}{16\pi^2}\frac{1}{(x_3)^2}\sum_{j}\delta_{s_j=s_{j+1}}\mathcal{O}^{s_1\dots s_j\, s_{j+1} \dots i_L}\tr(\scalc_{s_1}\dots\scalc_{s_{j-1}} \scalc_{s_{j+2}}\dots\scalc_{s_L})(x)\nonumber\\
 &\quad+\frac{\lambda}{8\pi^2} 
 \left(
 -\frac{1}{2 \peps} - \frac{1}{2}\log(4\pi)+\frac{1}{2}\gammaE - \log (x_3)+\digamma(\tfrac{k+1}{2})
 \right)
\frac{\Delta_{\text{1-loop}}}{2}\langle\mathcal{O}\rangle_{\text{tree}}(x)\eqndot 
 \end{align}
As $\mathcal{Z}_{\text{1-loop}}=\frac{\lambda}{16\pi^2}\frac{\Delta_{\text{1-loop}}}{2\peps}$ when using minimal subtraction, we have 
\begin{equation}
\label{eq: one-loop renormalisation constant contribution SU(2)}
 \langle\mathcal{O}\rangle_{\text{1-loop},\mathcal{Z}}(x)=\frac{\lambda}{16\pi}\frac{\Delta_{\text{1-loop}}}{2\peps}\langle\mathcal{O}\rangle_{\text{tree}}(x)\eqndot
\end{equation}
Thus, this contribution cancels the divergence above.%
\footnote{When using modified minimal subtraction, $\mathcal{Z}_{\text{1-loop}}=\frac{\lambda}{16\pi^2}\frac{\Delta_{\text{1-loop}}}{2\peps}\e^{-\peps\gammaE}(4\pi)^\peps$ and also the $- \frac{1}{2}\log(4\pi)+\frac{1}{2}\gammaE$ is cancelled.}
Moreover, the prefactor of $\log (x_3)\Delta_{\text{1-loop}}$ has the expected form coming from the one-loop correction to the scaling dimension.

The two-loop eigenstates are also known and can 
be efficiently obtained using one of the two recently developed 
technologies~\cite{Gromov:2012vu,Gromov:2012uv} and~\cite{Bargheer:2009xy,Jiang:2014mja}, both of which
build on the manipulation of an inhomogeneous version of the Heisenberg spin chain.
Hence, it only remains to calculate two overlaps, one involving a matrix-product state and 
an amputated one-loop Bethe state, and the other one involving a matrix product state and a two-loop correction to a Bethe state. 
These calculations should be doable~\cite{Work-in-Progress} adapting the technique  developed 
in~\cite{deLeeuw:2015hxa,Buhl-Mortensen:2015gfd}.

\subsection{One-loop one-point functions of \texorpdfstring{$\tr(Z^L)$}{tr(Z**L)}\label{sec:TrZL}}

Finally, let us consider the special case of the BPS operator $\tr(Z^L)$, i.e.\ $\mathcal{O}^{i_1 \dots i_L}=\prod_{j=1}^L(\delta_{i_j=3}+i\delta_{i_j=6})$.

At tree level, we have \cite{deLeeuw:2015hxa}
\begin{equation}\label{eq:tree}
\begin{aligned} 
 \langle\tr(Z^L)\rangle_{\text{tree}}(x) &=\frac{(-1)^L}{x_3^L}\sum_{i=1}^kd_{k,i}^L 
& =\begin{cases}
0\eqncom & L \text{ odd}\eqncom\\
-\frac{2}{x_3^{L}(L+1)}B_{L+1}\left(\frac{1-k}{2}\right)\eqncom & L \text{ even}\eqncom
\end{cases}
\end{aligned}
\end{equation}
where $d_{k,i}$ given in \eqref{eq: c and d coefficients} denotes the diagonal entries of $t_3$ and $B_{L+1}(u)$ is the Bernoulli polynomial of degree $L+1$.

The one-loop contributions $\langle\mathcal{O}\rangle_{\text{1-loop},\mathcal{Z}}(x)$ and $\langle\mathcal{O}\rangle_{\text{1-loop},\mathcal{O}}(x)$ vanish for this operator,
and \eqref{eq: general tadpole result} reduces to 
\begin{equation}
\label{eq: special tadpole result}
\begin{aligned}
 \langle\tr(Z^L)\rangle_{\text{1-loop,tad}}(x)=&L\tr((\scalc_{3})^{L-2}E^n{}_aE^a{}_{n'})(x)\left(\langle[\scalq_{3}]_{n,a}[\scalq_{3}]_{a,n'}\rangle-\langle[\scalq_{6}]_{n,a}[\scalq_{6}]_{a,n'}\rangle\right)\\
& +(k\times k)\text{-contributions} \eqncom
\end{aligned}
\end{equation}
where we have suppressed the argument $x$ of both propagators and the trivial summation over $j$ has produced a factor $L$. Inserting \eqref{eq:PropHardOffDiag} and \eqref{eq: propagator off-diagonal easy scalars}, the summation over $a$ produces a factor $(N-k)$ and the summation over $n,n'$ reduces the matrix unities to a unit matrix. 
Thus, we find%
\footnote{Recall that the lollipop contribution vanishes for all operators, cf.~\eqref{eq: correction to scalar vev main text}.}
\begin{equation}
\label{eq: special tadpole final result} 
 \langle\tr(Z^L)\rangle_{\text{1-loop}}(x)=\langle\tr(Z^L)\rangle_{\text{1-loop,tad}}(x)=\langle\tr(Z^{L-2})\rangle_{\text{tree}}(x)\frac{1}{x_3^2}\frac{\lambda}{16\pi^2}L+O(\tfrac{1}{N}) \eqndot
\end{equation}

\subsection{Finite-\texorpdfstring{$N$}{N} results}

In order to check our formalism and results, we have also computed the one-point functions explicitly in colour components for small $N,k$ using \texttt{Mathematica}. In this way, we explicitly diagonalised the mass matrix and used the mass eigenstates to find the propagators in colour space. We find that the mass spectrum perfectly matches \eqref{spectrum}. Moreover, from our explicit results for $N,k<9$, we were able to extract closed formulas for the one-point functions for any $N,k$. We find that they agree with \eqref{eq: general lollipop diagram final} and \eqref{eq: special tadpole final result} in the large-$N$ limit. The cancellations of divergencies for small mass, the regulator and irrational terms like $\gammaE$ all provide non-trivial consistency checks of our approach.

\paragraph{One-loop correction to vev}

From computations for $N,k<9$, we were able to find a closed expression for the vev of the scalar fields. In particular, our explicit computations show that the planar result
\begin{align}
\langle \scal_i \rangle_{\text{1-loop}} = 0
\end{align}
is actually exact. 

\paragraph{Tadpole correction to \texorpdfstring{$\tr (Z^L)$}{tr(Z**L)}}

Similarly, we have explicitly checked the tadpole diagrams for $N,k<9$. Again, we were able to find an exact expression for any $N,k,L$. It is given by
\begin{align}\label{eq:FullTadpole}
 \langle\tr(Z^L)\rangle_{\text{1-loop,tad}}(x) = L\frac{g^2_\YM}{8\pi^2}\frac{1}{x_3^L} \Bigg\{ &\frac{B_{L-1}\Big(\frac{k+1}{2}\Big)}{1-L} \left[N-k +\frac{k-1}{k} \frac{L-1}{2}\right] \nonumber\\ 
 &+\sum_{i=0}^{\lfloor\frac{k-2}{2}\rfloor} (H_{k-i-1} - H_i) \left[\frac{k-2i-1}{2}\right]^{L-1}\bigg\},
\end{align}
where $H_n = \sum_{i=1}^n i^{-1}$ are the harmonic numbers. Notice that \eqref{eq:FullTadpole} reduces to \eqref{eq: special tadpole final result} in the large-$N$ limit.

\section{Comparison to string theory for \texorpdfstring{$\langle \tr(Z^L)\rangle$}{<tr(Z**L)>}\label{stringtheory}}

When we wish to compare our perturbative, planar gauge-theory results to string theory, we are of course facing the eternal problem (and virtue) of the AdS/CFT correspondence that it is a strong-weak coupling duality. A proposal for how to circumvent this issue in the present set-up was put forward by Nagasaki, Tanida and Yamaguchi~\cite{Nagasaki:2011ue}. They pointed out that, compared to the usual AdS/CFT scenario, we here have at our disposal one extra tunable parameter, namely $k$, which 
 plays the role of the background gauge-field flux in the
string-theory picture and corresponds to the dimension of the $\SU2$ representation associated with the classical fields around which we expand on the gauge-theory side.
Hence, one can consider the double-scaling limit
\begin{equation}
\lambda\rightarrow \infty,\hspace{0.5cm}
k\rightarrow \infty, \hspace{0.5cm} \lambda/k^2\hspace{0.5cm} \mbox{finite},
\end{equation}
and furthermore consider $\lambda/k^2$ to be small.
The limit $\lambda\rightarrow \infty$ justifies a supergravity approximation on the string-theory side, whereas the 
assumption of $\lambda/k^2$ being small might bring one to the realm of perturbation theory for the field theory. This, however, requires that the gauge-theory perturbation series
for the observables of interest organises into an expansion in powers of $\lambda/k^2$. This idea is analogous to the BMN construction~\cite{Berenstein:2002jq}, where 
another large quantum quantum number, $J$, with the interpretation of an angular momentum, was considered to be large and was combined with $\lambda$ to form the double-scaling parameter $\lambda/J^2$. In the study of the spectral problem of ${\cal N}=4$ SYM theory, it was found that the perturbative expansion ceased to be an expansion in the parameter
$\lambda/J^2$ at four loops~\cite{Bern:2006ew,Beisert:2006ez,Cachazo:2006az}.

In~\cite{Nagasaki:2011ue}, the authors calculated in a supergravity approximation the one-point function of a special chiral primary of even length $L$, namely the unique one which carries $\SO{3}\times \SO{3}$ symmetry:
\begin{equation}\label{eq:O-Nagasaki}
{\cal O} (x)= C_L \tr \,\left(\left(\sum_{i=1}^3\phi_i^2\right)^{L/2}+\left(\sum_{i=4}^6\phi_i^2\right)
Q_{L-2}\left(\sum_{i=1}^3\phi_i^2,\sum_{i=4}^6\phi_i^2\right) 
\right)(x)\eqncom
\end{equation}
where $C_L$ is a normalisation constant and $Q_{L-2}(y,z)$ is a homogeneous polynomial of degree $\frac{L-2}{2}$ in 
$y$ and $z$.
This was done by considering the bulk-to-boundary propagator carrying the quantum numbers characteristic of the chiral primary, fixing one of its endpoints  to the point $x$ 
in the AdS boundary and integrating the other one over all points belonging to the D5-brane in the interior of $AdS_5\times S^5$. We note in passing that the computation can
be considerably simplified, not necessitating any integration, if one is only interested in the leading large-$L$ behaviour~\cite{Buhl-Mortensen:2015gfd}. However, we will include finite-$L$ corrections in the following discussion.  
The result for the string-theory one-point function found in~\cite{Nagasaki:2012re} turned out to be expandable as a series in the double-scaling parameter $\lambda/k^2$ and the leading term
in this expansion was shown to agree with the result of a tree-level computation in the gauge theory, which simply amounts to inserting the classical value for the fields 
into~(\ref{eq:O-Nagasaki}). 
The string-theory result of~\cite{Nagasaki:2012re} also implies a prediction for the gauge-theory result for the one-point function of the
operator above at next-to-leading order in the double-scaling  parameter. The chiral primary~(\ref{eq:O-Nagasaki}) differs from the one we focused on in section~\ref{sec:TrZL},
namely $\tr(Z^L)$,  but one can easily convince oneself that the latter has a non-vanishing projection on the former. This implies that the ratio between the next-to-leading-order
contribution and the leading-order contribution in $\lambda/k^2$ should be the same for the two operators. The prediction for this ratio following from the
analysis of~\cite{Nagasaki:2012re} reads
\begin{equation}
\left.\frac{\langle {\cal O}\rangle_{\text{1-loop}}}{\langle {\cal O}\rangle_{\text{tree-level}}}\right|_{\text{string}}=
\frac{\lambda}{4\pi^2 k^2} \frac{L(L+1)}{L-1}\eqndot
\end{equation}

Combining \eqref{eq:tree} and \eqref{eq: special tadpole final result},
we likewise have a result for this quantity:
\begin{equation}
\left.\frac{\langle {\cal O}\rangle_{\text{1-loop}}}{\langle {\cal O}\rangle_{\text{tree-level}}}\right|_{\text{gauge}}=
\frac{\lambda}{4\pi^2 k^2}\left( \frac{L(L+1)}{L-1} + O(k^{-2})\right)\eqncom
\end{equation}
which perfectly matches the string-theory prediction.
This constitutes a highly nontrivial test of the AdS/dCFT correspondence! 
Whether the field theory result continues to organise into a power series expansion in the double-scaling parameter $\lambda/k^2$ at higher loop order is obviously a question which requires further investigation. As already mentioned, the BMN expansion broke down at four-loop order. Nevertheless, the BMN idea was instrumental in catalysing the integrability approach to AdS/CFT. One could dream that the present double-scaling 
idea would play a similarly instrumental role for the study of AdS/dCFT.

\section{Conclusion and outlook}
\label{sec: conclusion and outlook}

With the present paper, we have performed a non-trivial, positive test of the gauge-gravity correspondence in a set-up where both the supersymmetry and the conformal symmetry
are partially broken. 
In order to carry out the test, we had to set up the framework for loop computations in 
a Higgsed defect version of ${\cal N}=4$ SYM theory, dual to a D5-D3 probe brane system with flux.
This framework now opens the possibility of calculating a large amount of observables of the theory and hence obtaining more insight into the properties of the AdS/dCFT setup in general and the specific dCFT in particular. 
As an application, we formulated the precise line of action  
for calculating the one-loop correction to any scalar operator,  leaving only a combinatorial problem that should be solvable invoking the tools of integrability.
In particular, we have found that only two Feynman diagrams are relevant for the calculation
and we have evaluated these using dimensional regularisation finding that one of them vanishes.
So far, we have completed the calculation of  the one-loop correction to the one-point function of the BMN vacuum which 
we previously summarised in~\cite{Buhl-Mortensen:2016pxs}.  For this particular correlator, a comparison with string theory is possible in 
a certain double-scaling limit and a perfect match is found.
A similar situation occurs in a calculation of the expectation value of a straight Wilson line~\cite{deLeeuw:2016vgp}.  

Apart from the two simple observables just mentioned, there exist at the time of writing no other string-theory results that one could compare to and it would be interesting and important to
extend the string-theory computations to other cases. The most immediate one would be one-point functions of spinning strings
corresponding to non-protected operators of the $\SU2$ subsector. 

One-point functions only constitute one out of several novel types of correlators specific to dCFTs. Another class of such operators are two-point functions between operators with different conformal dimensions.  General arguments 
constrain the space-time dependence of such two point functions~\cite{Cardy:1984bb} 
 and it would be interesting  to
demonstrate by explicit computation that the constraints are met both from the particular dCFT considered here and from its string-theory counterpart.

Until now, we have focused on one-loop computations for which the defect fields do not play any role. A natural
new direction of investigation would be to consider situations where the defect fields come into play. We expect that this will
happen if the present calculation is carried on to higher-loop order.
Defect fields can of course also appear in 
correlation functions either with other defect fields or with bulk fields. Correlation functions between defect and bulk fields
again constitute a novel type of observables for which only very few explicit results are known~\cite{DeWolfe:2001pq}.

The D5-D3 probe brane set-up is only one out of a number of probe brane set-ups which have dual dCFTs, see for instance~\cite{Ammon:2015wua}. 
Another set-up which is very reminiscent of
the one considered here is the D7-D3 probe brane system where the geometry of the D7 brane is either  $AdS_4\times S^4$ or $AdS_4\times S^2\times S^2$ and where again
a certain background gauge field has a non-vanishing flux through either $S^4$ or $S^2\times S^2$, making possible the definition of a double-scaling parameter.
The dual dCFT is again a defect version of ${\cal N}=4$ SYM theory but the
set-up is no longer supersymmetric. So far,  for this dCFT  only tree-level one-point functions of chiral primaries have been calculated and these were found to match a string-theory
prediction to the leading order in the double-scaling parameter~\cite{Kristjansen:2012tn}. It would be interesting to extend this study to non-protected operators~\cite{deLeeuw:2016ofj}
as well as to generalise the approach presented in this paper to proceed to one-loop order. The latter endeavour, however, 
is likely to involve novel complications and subtleties due to the complete absence of supersymmetry.

 The development of the last 15 years has lead to numerous discoveries of novel features of 
  ${\cal N}=4$ SYM theory and the AdS/CFT correspondence as well as novel techniques applicable to this set-up, such as
 integrability~\cite{Beisert:2010jr}, localisation~\cite{Zarembo:2016bbk}, the conformal bootstrap~\cite{Beem:2013qxa} and the duality between Wilson loops  and correlators~\cite{Alday:2010zy}.   
The tools of integrability have already proven useful in the present set-up, in particular at tree level where they permitted the derivation of a close form for the one-point function valid for any operator in the $\SU{2}$ subsector and for any value of the parameter $k$~\cite{Buhl-Mortensen:2015gfd,deLeeuw:2015hxa}, but also for the present one-loop
considerations where they come into play for instance in section~\ref{sec:SU(2)}. Whether integrability tools will facilitate going to higher loop orders or to other subsectors remains to be seen. 
A generalisation of the conformal bootstrap approach to the defect set-up has been studied
in~\cite{Liendo:2012hy,Gliozzi:2015qsa,Billo:2016cpy,Liendo:2016ymz}. 
It would be interesting to investigate in more detail how far  this as well as the other above mentioned techniques can be taken in the context of the present dCFT.

\section*{Acknowledgements}

We thank S.\ Caron-Huot,  G.\ Korchemsky, C.\ Sieg, and in particular G.\ Semenoff and K.\ Zarembo for useful discussions. 
I.B.-M.,  M.d.L., A.C.I., C.K.\  and M.W.\ were  supported  in  part  by  FNU  through
grants number DFF-1323-00082 and DFF-4002-00037. A.C.I.\ in addition was supported by  the ERC Advanced Grant 291092.
All authors acknowledge the kind hospitality of NORDITA during the program ``Holography and Dualities.''  M.d.L., A.C.I., C.K.\  and M.W.\
in addition acknowledge the kind hospitality of Humboldt University
during the KOSMOS summer university program ``Integrability for the Holographic Universe.''

\appendix

\section{Explicit form of the representation matrices\label{representationmatrices}}

We present here explicit expressions for the representation matrices $t_i$ in the $k$-dimensional irreducible representation of the Lie algebra $\SU{2}$.

Following \cite{deLeeuw:2015hxa}, we define the standard matrices $E^i{}_j$ satisfying 
\begin{equation}
 E^i{}_jE^k{}_l=\delta^k{}_jE^i{}_l\eqndot
 \label{eq:E-E-mult}
\end{equation}
We define 
\begin{equation}
\label{eq:SU2byEmats}
 t_+=\sum_{i=1}^{k-1}c_{k,i}E^i{}_{i+1}\eqncom \qquad t_-=\sum_{i=1}^{k-1}c_{k,i}E^{i+1}{}_{i}\eqncom \qquad t_3=\sum_{i=1}^{k}d_{k,i}E^i{}_{i}\eqncom
\end{equation}
where 
\begin{equation}
\label{eq: c and d coefficients}
 c_{k,i}=\sqrt{i(k-i)}\eqncom \qquad d_{k,i}=\frac{1}{2}(k-2i+1)\eqndot
\end{equation}
The standard $k$-dimensional representation of the Lie algebra $\SU{2}$ is then given by 
\begin{equation}
 t_1=\frac{t_++t_-}{2}\eqncom \qquad t_2=\frac{t_+-t_-}{2i}\qquad\text{and} \qquad t_3\eqndot
\end{equation}

\section{`Spherical' colour basis and the fuzzy sphere\label{fuzzy}}

In this appendix, we summarise some properties of the spherical harmonics of the fuzzy sphere, which are used in the diagonalisation of the mass matrix in section \ref{diagonalisation}.

Let $\Phi$ be any adjoint field. It transforms naturally under $\SU2$ as
\begin{equation}
  \Phi \to \e^{-i\lambda_i t_i} \Phi \e^{i\lambda_i t_i},
\end{equation}
or infinitesimally 
\begin{equation}
  \delta\Phi  =  -i\lambda_i \Ad(t_i) \Phi = -i\lambda_i [t_i,\Phi]\eqndot
\end{equation}
As usual, we can decompose this representation into a sum of irreducible 
representations. To do this explicitly for the components in the $k\times k$ block, we use the spherical harmonics
$Y_\ell^m$; see \cite{Hoppe82,deWit:1988wri}. We start by remembering that $r^\ell Y_\ell^m$
can be written as
a homogeneous polynomial of order $\ell$ in the Cartesian coordinates.
In detail, we have
\begin{equation}
  r^\ell Y_\ell^m = (-1)^m \sqrt{2\ell+1} \bar{\Pi}_\ell^m (x_1+ix_2)^m\eqncom
  \qquad r^\ell Y_\ell^{-m} =  \sqrt{2\ell+1} \bar{\Pi}_\ell^m (x_1-ix_2)^m\eqncom
\end{equation}
for $m \geq 0$ and with
\begin{equation}
  \bar{\Pi}_\ell^m
    = \sqrt{\frac{(\ell-m)!}{(\ell+m)!}}\sum_{s=0}^{\lfloor (\ell-m)/2 \rfloor}
      (-1)^s 2^{-\ell}\binom{\ell}{s}\binom{2\ell-2s}{\ell}
      \frac{(\ell-2s)!}{(\ell-2s-m)!}r^{2s}x_3^{\ell-2s-m}\eqndot
\end{equation}
Note that $x_1,x_2,x_3$ have nothing to do with the physical coordinates.
It follows that there is a symmetric set of coefficients $f^{\ell m}_{i_1,i_2,\ldots i_\ell}$
such that
\begin{equation}
  r^\ell Y_\ell^m = \sum_{\{i\}} f^{\ell m}_{i_1,i_2,\ldots i_\ell}
    x_{i_1}\cdots x_{i_\ell}\eqndot
  \label{eq:harmonic-polynomial}
\end{equation}

We now want to define a $N\times N$ matrix corresponding to $Y_\ell^m$.
We rescale the $\SU{2}$ generators to 
\begin{equation}
  \hat{x}_i = \sqrt{\frac{4}{k^2-1}} t_i\eqndot
\end{equation}
These are coordinates on the fuzzy unit sphere. In particular, we have
\begin{equation}
  \hat{x}^2 = \hat{x}_i \hat{x}_i = 1
\end{equation}
as an operator identity. Substituting these operators into \eqref{eq:harmonic-polynomial},
we obtain the operators%
\footnote{Note that for $\ell \geq k$ this construction simply gives zero.}
\begin{equation}
  \tilde{Y}_\ell^m = \sum_{\{i\}} f^{\ell m}_{i_1,i_2,\ldots i_\ell}
    \hat{x}_{i_1}\cdots \hat{x}_{i_\ell}\eqncom\qquad
    \ell = 1,\ldots,k-1\eqndot
\end{equation}
These operators achieve the decomposition of the $\SU{2}$ representation \eqref{eq: rewriting k x k} in the $k\times k$ block, cf.\ \cite{Hoppe82,deWit:1988wri}. 
In particular, they satisfy \eqref{eq:fuzzy-eigenvalues}.

The $\tilde{Y}_\ell^m$ form a orthogonal basis for the traceless $k\times k$
matrices, but they are not normalised. If we define\footnote{The normalisation
  constant follows from~\cite{Hoppe82}.
  }
\begin{equation}
  \hat{Y}_\ell^m = \sqrt{\frac{(k-\ell-1)!}{(k+\ell)!}} 2^\ell 
    \left(\frac{k^2-1}{4}\right)^{\ell/2} \tilde{Y}_\ell^m\eqncom
\end{equation}
we have
\begin{equation}\label{eq:Ynorm}
  \tr[(\hat{Y}_\ell^m)^\dagger \hat{Y}_{\ell'}^{m'}]
    = \delta_{\ell\ell'}\delta_{mm'}\eqncom \quad \text{where}\quad (\hat Y_l^m)^\dagger = (-1)^m \hat Y_l^{-m}\eqncom
\end{equation}
and thus
\begin{equation}
 \label{eq:Ynorm used}
  \tr[\hat{Y}_\ell^m \hat{Y}_{\ell'}^{m'}]
    = (-1)^m\delta_{\ell\ell'}\delta_{m+m',0}\eqndot
\end{equation}

The matrix elements of the fuzzy spherical harmonics can be found in \cite{Kawamoto:2015qla} up to normalisation; we normalise them to satisfy \eqref{eq:Ynorm}. They are given explicitly by
\begin{equation}
\label{eq: matrix element Y}
[\hat Y^m_\ell]_{n,n'} = (-1)^{k-n}\sqrt{2\ell +1}\mmatrix[1][10]{\frac{k-1}{2} & \ell & \frac{k-1}{2} \\ n-\frac{k+1}{2} & m  & -n'+\frac{k+1}{2} }\eqncom \qquad n,n' = 1, \ldots, k\eqncom
\end{equation}
where the large parenthesis denote Wigner's 3$j$ symbol.
Hence,
\begin{equation}
 \hat Y^m_\ell=[\hat Y^m_\ell]_{n,n'}E^n{}_{n'}\eqndot
\end{equation}
Inverting this equation using the orthogonality and normalisation of  $\hat Y^m_\ell$ and $E^n{}_{n'}$, we find
\begin{equation}
 E^n{}_{n'} =[\hat Y^m_\ell]_{n,n'}\hat Y^m_\ell\eqndot
\end{equation}

Note that $\hat Y^m_\ell$ transforms in the spin-$\ell$ representation under $L_i$, i.e.
\begin{equation}
  L_i \hat Y^m_\ell = [t_i^{(k)},\hat Y^m_\ell] = \hat Y^{m'}_\ell [t_i^{(2\ell+1)}]_{\ell-m'+1,\ell-m+1}\eqncom
\end{equation}
where $t_i^{(k)}\equiv t_i$ denotes the generators of the $k$-dimensional irreducible representation given in appendix \ref{representationmatrices} and $t_i^{(2\ell+1)}$ denotes the analogous generators of the $(2\ell+1)$-dimensional irreducible representation.

Finally, for $\ell=1$ the spherical harmonics can be explicitly related to our $t_i$ matrices:
\begin{equation}\label{eq:TviaY}
\begin{aligned}
t_1 &= \frac{(-1)^{k+1}}{2}\sqrt{\frac{k(k^2-1)}{6}} (\hat{Y}^{-1}_1 -\hat{Y}^1_1)\eqncom\\
t_2 &= i\frac{(-1)^{k+1}}{2}\sqrt{\frac{k(k^2-1)}{6}} (\hat{Y}^{-1}_1 + \hat{Y}^1_1 )\eqncom\\
t_3 &= \frac{(-1)^{k+1}}{2}\sqrt{\frac{k(k^2-1)}{3}} \hat{Y}^0_1\eqndot 
\end{aligned}
\end{equation}

\section{Decomposition of 10-D Majorana-Weyl fermions\label{app: 10Dto4D}}

In this appendix, we present our conventions for the decomposition of the ten-dimensional fermion into the four-dimensional fermions and the corresponding gamma matrices.

The ten-dimensional Majorana-Weyl fermions satisfy
\begin{equation}
 \tenDferm = \mc C_{10}\bar\tenDferm^T\eqncom \qquad \Gamma^{11}\tenDferm = -\tenDferm\eqncom \label{eq:Majorana-Weyl}
\end{equation}
where $\Gamma^M$ are ten-dimensional gamma matrices satisfying%
\footnote{Recall that we are using mostly-positive signature.}
\begin{equation}
 \{\Gamma^M, \Gamma^N\} = -2 \eta^{MN}\eqndot
\end{equation}
We proceed to decompose the ten-dimensional gamma matrices in term of four-dimensional ones. The four-dimensional gamma matrices are $\gamma^\mu$, $\mu = 0,1,2,3$, and we choose the representation
\begin{equation}
\gamma^\mu = \mmatrix{ 0 & \sigma^\mu \\ \bar \sigma^\mu & 0 }\eqncom \qquad \{\gamma^\mu, \gamma^\nu\} = -2\eta^{\mu\nu}\eqncom
\end{equation}
where $\sigma^\mu=(\idm_2,\sigma^i)$ and $\bar\sigma^\mu=(\idm_2,-\sigma^i)$.
We also have
\begin{equation}
\gamma^5 = i \gamma^0\gamma^1\gamma^2\gamma^3
\end{equation}
and the charge conjugation matrix
\begin{equation}
\mc C = \mmatrix{0 & 1 & 0 & 0 \\ -1 & 0 & 0 & 0 \\ 0 & 0 & 0 & -1 \\ 0 & 0 & 1 & 0 }, \qquad \gamma_\mu^T = -\mc C \gamma_\mu \mc C^{-1}\eqndot
\label{eq:charge-conjugation-matrix}
\end{equation}
It follows that a Lorentz invariant reality condition is
\begin{equation}
\psi = \psi^C, \qquad \psi^C \equiv \mc C \bar \psi^T\eqncom \label{eq:Majorana}
\end{equation}
where $\bar \psi = \psi^\dagger \gamma_0$.

We adopt the following representation for the ten-dimensional Clifford algebra
\begin{align}
\Gamma^\mu &= \gamma^\mu \otimes \mds 1_8 \eqncom  & \mu &= 0,1,2,3,\\
\Gamma^{i+3} &
=  \tilde{\Gamma}^i = \gamma^5 \otimes\begin{pmatrix}0 & -G^{i} \\ G^{i} & 0 \end{pmatrix}\eqncom & i &= 1,2, 3, \\
\Gamma^{i+3} &
=  \tilde{\Gamma}^i = \gamma^5 \otimes\begin{pmatrix}0 & G^{i} \\ G^{i} & 0 \end{pmatrix}\eqncom & i &= 4,5,6,
\end{align}
where $G^i$ are the $4 \times 4$ matrices
\begin{equation}
 \label{eq:G123}
\begin{aligned}
&G^1 = i
\begin{pmatrix}
0 & -\sigma_3  \\
\sigma_3 & 0 
\end{pmatrix} \eqncom 
&&G^2 =i
\begin{pmatrix}
 0 & \sigma_1 \\
 -\sigma_1 & 0 \\
\end{pmatrix}
\eqncom 
&&G^3 = 
\begin{pmatrix}
 \sigma_2 & 0 \\
 0 & \sigma_2 
\end{pmatrix}
\eqncom\\
&G^4 = i
\begin{pmatrix}
 0 & -\sigma_2 \\
 -\sigma_2 & 0 
\end{pmatrix}
\eqncom
&&G^5 = 
\begin{pmatrix}
 0 & -\idm_2 \\
 \idm_2 & 0 
\end{pmatrix}
\eqncom
&&G^6 = i
\begin{pmatrix}
 \sigma_2 & 0 \\
 0 & -\sigma_2
\end{pmatrix}
\eqndot
\end{aligned}
\end{equation}
The latter satisfy
\begin{align}\{G^i, G^j\} &= \left\{\begin{array}{lr} +2\delta^{i,j},\qquad & \qquad\qquad\qquad i,j = 1,2,3, \\ -2\delta^{i,j}, \qquad&\qquad\qquad\qquad i,j = 4,5,6, \end{array}\right. \label{eq:G-anti-commutator}\\
[G^i, G^j] &= \left\{\begin{array}{lr} -2 i\,  \teps^{ijk}G^k, & i,j = 1,2,3, \\ + 2  \, \teps^{ijk}G^k, &\,\, i,j = 4,5,6, \\ 0,  &   i = 1,2,3, \quad j = 4,5,6. \end{array}\right. 
\end{align}
Finally, the ten-dimensional charge conjugation matrix and $\Gamma_{11}$ are given by
\begin{equation}
\mc C_{10} = \mc C \otimes \mmatrix{0 & \mds 1_4 \\ \mds 1_4 & 0 }, \qquad \Gamma_{11} = \gamma_5 \otimes \mmatrix{-\mds 1_4 & 0 \\ 0 & \mds 1_4}\eqndot
\end{equation}
Imposing the Majorana-Weyl constraint (\ref{eq:Majorana-Weyl}) on a ten-dimensional fermion is now seen to imply
\begin{equation}
\tenDferm = \mmatrix{L\psi_1 \\ \vdots \\ L\psi_4\\ R \psi_1 \\ \vdots \\ R \psi_4 }\eqncom
\end{equation}
where
\begin{equation}
L = \half( \mds 1 + \gamma_5), \qquad R = \half(\mds 1 - \gamma_5)
\end{equation}
act on four-dimensional Majorana fermions $\psi_i$ satisfying (\ref{eq:Majorana}).

Using the above decomposition of the ten-dimensional fermions and gamma matrices, we find
\begin{align} \frac{1}{2}\bar \tenDferm_j \tilde\Gamma^i_{jk}[\phi_i, \tenDferm_k] 
&= \frac{1}{2}\sum_{i =1}^3\bar\psi_jG^i_{jk}[\phi_i, \psi_k] + \frac{1}{2}\sum_{i=4}^6\bar\psi_jG^i_{jk}[\phi_i, \gamma_5\psi_k] \eqncom
\end{align}
and hence the fermion mass term reads
\begin{equation}
	-\frac{1}{2x_3}\sum_{i =1}^3\bar\psi_jG^i_{jk}[t_{i}, \psi_k]\eqndot 
\end{equation}

\section{One-loop correction to the scalar vevs}
\label{app: one-loop vevs}

In this appendix, we compute the one-loop correction to the vevs of the scalar fields. To this loop order, we only need to take cubic vertices into account as only diagrams of lollipop type contribute. 
The one-loop correction takes the form
\begin{equation}\label{eq:LollipopD}
\langle\phi_i \rangle_{\text{1-loop}}(x)= 
\contraction{}{\scalq}{_{i}(x)
\int \de^4 y \sum_{\Phi_1,\Phi_2,\Phi_3}V_3(}{\Phi}\scalq_{i}(x)
\int \de^4 y \sum_{\Phi_1,\Phi_2,\Phi_3}V_3(\Phi_1(y),\contraction{}{\Phi}{_2(y),}{\Phi}\Phi_2(y),\Phi_3(y))\eqndot
\end{equation}
There are three parts to the computation of the above vev: the contractions of the fields in the vertex, the integral and the external contraction corresponding to the stick of the lollipop.
However, we will see that the sum of all the contractions in the vertex already vanishes after partial integration, and thus
\begin{equation}
 \langle\phi_i \rangle_{\text{1-loop}}(x)= 0\eqndot
\end{equation}
Moreover, the one-loop corrections to the vevs of all other individual fields also vanish. 

 \subsection{Contractions of the fields in the loop}
 \label{subsec: loop}

From the cubic interaction terms in the action \eqref{eq:cubic} and the form of the propagators in section \ref{subsec: mixing part of propagators}, we find the externally contracted field in the vertex can be either $\Phi_1=\scalq_i$ or $\Phi_1=A_\mu$.%
\footnote{We have no non-vanishing contraction for $\Phi_1=\ferm$, which would lead to a potentially non-vanishing vev of a single fermion.}
There are then three possible types of loops. We can have easy bosons $E$ and ghosts, complicated bosons $C$ or fermions running in the loop. When we evaluate the loop, all the propagators are taken at the same point $y$ in space-time. Moreover, we will also work in the planar limit. 

\paragraph{Contribution of easy scalars, easy gauge fields and ghosts in the loop}

Let us first consider the contribution of easy scalars, easy gauge fields and ghosts running in the loop of the lollipop diagrams, where we restrict ourselves to the off-diagonal $k\times(N-k)$ and $(N-k)\times k$ blocks that contribute in the large-$N$ limit. 

We start with diagrams for which $\Phi_1=\scalq_i$.
For the sake of concreteness, we focus on the easy scalar $\scalq_4$ running in the loop; the contributions of all other easy fields are essentially the same. The corresponding interaction term is \eqref{eq:cubic} 
\begin{equation}
 +\tr([\scalc_i,\scalq_4][\scalq_i,\scalq_4]) = +\tr(\scalq_i[\scalq_4,[\scalc_i,\scalq_4]])=-\frac{1}{y_3}\tr(\scalq_i[\scalq_4,[t_i,\scalq_4]])\eqndot
\end{equation}
From the decomposition \eqref{eq: decomposition into blocks} of $\scalq_4$, we find  
\begin{equation}
 \tr(\scalq_i[\contraction{}{\scalq}{_4,[t_i,}{\scalq}\scalq_4,[t_i,\scalq_4]])\simeq -\langle[\scalq_4]_{n,a}[\scalq_4]_{a,n'}\rangle\left(
 \tr(\scalq_iE^n{}_{n'}t_i)+\tr(\scalq_it_iE^n{}_{n'})\right)
 \eqncom
\end{equation}
where we have dropped the contributions from the components in the $k\times k$ block, which are irrelevant in the large-$N$ limit. We denote the restriction to terms relevant in the large-$N$ limit by $\simeq$.
Using the explicit form of the propagator \eqref{eq: propagator off-diagonal easy scalars}, 
the matrices $E^n{}_{n'}$ become unit matrices after the summation over $n,n'$, the $a$ summation yields a factor $N-k$ and we find in the large-$N$ limit
\begin{equation}
 +\tr(\scalq_i[\contraction{}{\scalq}{_4,[\scalc_i,}{\scalq}\scalq_4,[\scalc_i,\scalq_4]])\simeq  \frac{2N}{y_3}K^{m^2=\frac{k^2-1}{4}} \tr(\scalq_it_i)\eqndot
\end{equation}
In total, this contribution has a prefactor of $n_{\phi,\text{easy}}+n_{A,\text{easy}}-n_{c}$.

Let us now turn to the effective vertices that involve $\Phi_1=A_\mu$. We again focus on the easy scalar $\scalq_4$ running in the loop. The corresponding vertex is
\begin{align}
i\tr([A^\mu,\scalq_4]\partial_\mu\scalq_4) =i
\tr(A^\mu [\scalq_4,\partial_\mu\scalq_4]) \eqndot
\end{align}
We contract the scalar fields and obtain
\begin{align}
\label{eq: vanishing easy contraction with derivative}
i\tr(A^\mu [\contraction{}{\scalq}{_4,\partial_\mu}{\scalq}\scalq_4,\partial_\mu\scalq_4])
\simeq
i\big[
\langle [\scalq_4]_{n,a}\partial_\mu[\scalq_4]_{a,n^\prime}  \rangle -
i\langle \partial_\mu[\scalq_4]_{n,a}[\scalq_4]_{a,n^\prime} \rangle\big]
\tr(A^\mu E^n{}_{n^\prime})=0
\eqncom
\end{align}
where the last step follows from the symmetry of the propagator. Similarly, the contractions of 
\begin{align}
&i[A^\mu,A^\nu]\partial_\mu A_\nu\eqncom
&&i(\partial_\mu\bar{c})[A^\mu,c]
\end{align}
with the easy gauge fields and ghosts running in the loop are also vanishing.

\paragraph{Contribution from complicated bosons in the loop}

For the case of complicated bosons contracted in the loop, there are two vertices with insertions of the classical fields that can contribute:
\begin{align}\label{eq:Vertices Hard Eff Vertex}
 +\tr([\scalc_i,\scalq_j][\scalq_i,\scalq_j]) &= -\frac{1}{y_3}\tr(\scalq_i[\scalq_j,[t_i,\scalq_j]])\eqncom \nonumber\\
 +\tr([A^\mu, \scalc_i][A_\mu,\scalq_i]) &= -\frac{1}{y_3}\tr(\scalq_i[A^\mu,[t_i,A_\mu]])\eqndot
\end{align}
The requirement that the boson in the loop is complicated effectively fixes $i,j=1,2,3$ and $\mu=3$.

The fields at the vertex can be contracted in three different ways. Let us for simplicity restrict to the vertex with $\Phi_1=\phi_i$. We can connect $\scalq_j$ to $\scalq_j$ and there are two ways we can connect $\scalq_j$ to $\scalq_i$:
\begin{align}
&\contraction{\tr(\scalq_i[}{\scalq}{_j,[t_i,}{\scalq} \tr(\scalq_i[\scalq_j,[t_i,\scalq_j]]),
&&\contraction{\tr(}{\scalq}{_i[\scalq_j,[t_i,}{\scalq} \tr(\scalq_i[\scalq_j,[t_i,\scalq_j]]),
&&\contraction{\tr(}{\scalq}{_i[}{\scalq} \tr(\scalq_i[\scalq_j,[t_i,\scalq_j]])\eqndot
\end{align}
The terms with $A_3$ can be contracted analogously. 

Out of the above three contractions, the easiest one to compute is the first one. 
Again, we work in the planar limit and the computation is similar to the easy bosons discussed above. From \eqref{eq:PropHardOffDiag}, we then immediately find
\begin{equation}
\contraction{\tr(\scalq_i[}{\scalq}{_1,[t_i,}{\scalq} \tr(\scalq_i[\scalq_1,[t_i,\scalq_1]])
\simeq -N \!\left[\!\frac{k+1}{k}K^{m^2=\frac{(k-2)^2-1}{4}}+\frac{k-1}{k}K^{m^2=\frac{(k+2)^2-1}{4}}\!\right]\!\tr(\scalq_it_i)\eqndot
\end{equation}
From \eqref{eq:PropHardOffDiag}, it is easy to see that all the complicated bosons give the same contribution, which results in an overall factor of $n_{\phi,\complicated}+n_{A,\complicated}$. 

The other two contractions are more involved but share a similar structure. Let us work out the last one first. We obtain
\begin{align}
&\contraction{\tr(}{\scalq}{_i[}{\scalq} \tr(\scalq_i[\scalq_j,[t_i,\scalq_j]]) \simeq (\langle [\scalq_i]_{a,n}[\scalq_j]_{n^\prime,a}\rangle-\langle [\scalq_j]_{a,n}[\scalq_i]_{n^\prime,a}\rangle) \tr(E^{n}{}_{n^\prime}[t_i,\scalq_j]) \eqndot
\end{align}
Inserting the explicit form of the propagator \eqref{eq:PropHardOffDiag}, it is easy to see that the contribution of the term with $\delta_{n,n^\prime}$ cancels and we are left with
\begin{align}
\contraction{\tr(}{\scalq}{_i[}{\scalq} \tr(\scalq_i[\scalq_j,[t_i,\scalq_j]]) 
&\simeq  -2 i \frac{N}{k} \left( K^{m^2=\frac{(k-2)^2-1}{4}}-K^{m^2=\frac{(k+2)^2-1}{4}}\right)\teps_{ijk} \tr(t_k [t_i,\scalq_j]) \nonumber \\
&= 2 \frac{N}{k} \left(K^{m^2=\frac{(k-2)^2-1}{4}}-K^{m^2=\frac{(k+2)^2-1}{4}} \right) \teps_{ijk}\teps_{kil} \tr(t_l \scalq_j)\nonumber \\
&= 2(n_{\phi,\complicated} -1) \frac{N}{k} \left( K^{m^2=\frac{(k-2)^2-1}{4}}-K^{m^2=\frac{(k+2)^2-1}{4}} \right) \tr(t_i \scalq_j)\eqndot
\end{align}
The final contraction gives
\begin{align}
&\contraction{\tr(}{\scalq}{_i[\scalq_j,[t_i,}{\scalq} \tr(\scalq_i[\scalq_j,[t_i,\scalq_j]]) \simeq \langle [\scalq_i]_{a,n}[\scalq_j]_{n^\prime,a^\prime}\rangle \tr(E^{n}{}_{n^\prime} t_i\scalq_j)+ \langle [\scalq_j]_{n,a}[\scalq_i]_{a^\prime,n^\prime}\rangle \tr(E^{n}{}_{n^\prime} \scalq_j t_i) \eqndot
\end{align}
The second term in the propagator  \eqref{eq:PropHardOffDiag} evaluates in the same way as above, but the $\delta_{n,n^\prime}$ term now also contributes and we obtain
\begin{align}
\contraction{\tr(}{\scalq}{_i[\scalq_j,[t_i,}{\scalq} \tr(\scalq_i[\scalq_j,[t_i,\scalq_j]]) 
&\simeq N \left(\frac{k+1}{k} K^{m^2=\frac{(k-2)^2-1}{4}}+\frac{k-1}{k} K^{m^2=\frac{(k+2)^2-1}{4}}\right)\tr(\scalq_it_i) \nonumber \\
&\qquad + (n_{\phi,\complicated} -1) \frac{N}{k} (K^{m^2=\frac{(k-2)^2-1}{4}}-K^{m^2=\frac{(k+2)^2-1}{4}})\tr(\scalq_it_i) \eqndot
\end{align}

The vertices from \eqref{eq:Vertices Hard Eff Vertex} with $\Phi_1=A_3$ instead of $\Phi_1=\scalq_i$ contribute with 
\begin{align}
\contraction{\tr(}{\scalq}{_i[}{A} \tr(\scalq_i[A_3,[t_i,A_3]])  = \contraction{\tr(}{\scalq}{_i[A_3,[t_i,}{A} \tr(\scalq_i[A_3,[t_i,A_3]])  &\simeq 0\eqncom
\end{align}
as can be seem from a short analogous calculation.

Finally, there is a non-trivial contribution from the vertex
\begin{align}
\label{eq: partial derivative vertex}
\tr( i[A^\mu,\scalq_i]\partial_\mu\scalq_i)
\eqncom
\end{align}
which can be contracted non-trivially in two different ways that contribute for $\Phi_1=\phi_i$:
\begin{align}
&\contraction{i\tr([}{A^3}{,}{\scalq}\tr( i[A^3,\scalq_i]\partial_3\scalq_i)\eqncom
&&\contraction{i\tr([}{A^3}{,\scalq_i]\partial_3}{\scalq}\tr (i[A^3,\scalq_i]\partial_3\scalq_i)\eqndot
\end{align}
In the large-$N$ limit, the only terms that survive are 
\begin{align}
\contraction{i\tr([}{A^3}{,}{\scalq}\tr( i[A^3,\scalq_i]\partial_3\scalq_i)&\simeq 
2i  \langle [A^3]_{n,a}[\scalq_i]_{a,n^{\prime}} \rangle \tr (E^n{}_{n^\prime}\partial_3 \scalq_i )\nonumber\\
&\simeq2\frac{N}{k}\left(K^{m^2=\frac{(k-2)^2-1}{4}}-K^{m^2=\frac{(k+2)^2-1}{4}}\right) \tr ( t_i \partial_3 \scalq_i )
\end{align}
and
\begin{align}
\contraction{i\tr([}{A^3}{,\scalq_i]\partial_3}{\scalq}\tr (i[A^3,\scalq_i]\partial_3\scalq_i)&\simeq 
2i  \langle [\partial_3\scalq_i]_{n,a} [A^3]_{a,n^{\prime}} \rangle \tr (E^n{}_{n^\prime} \scalq_i )\nonumber\\
&\simeq-\frac{N}{k}\partial_3\left(K^{m^2=\frac{(k-2)^2-1}{4}}-K^{m^2=\frac{(k+2)^2-1}{4}}\right) \tr ( t_i \scalq_i )\eqndot
\end{align}
In the last line, we expressed the propagator with a derivative on the field as a derivative of the propagator. It follows from the identity
\begin{align}
\lim_{x\rightarrow y} \langle [A^3(x)]_{n,a}[\partial_3\scalq_i(y)]_{a,n^{\prime}} \rangle =
\half \partial_{y_3} \lim_{x\rightarrow y} \langle [A^3(x)]_{n,a}[\scalq_i(y)]_{a,n^{\prime}} \rangle  \eqncom
\end{align}
which follows from the explicit form of the propagator \eqref{eq:PropagatorBessel} and the following property of the Bessel functions
\begin{align}
\half\partial_x\Big[I_{\nu-1}(x)K_{\nu-1}(x)-I_{\nu+1}(x)K_{\nu+1}(x)\Big] = \Big[\partial_x I_{\nu-1}(x)\Big]K_{\nu-1}(x)-\Big[\partial_x I_{\nu+1}(x)\Big]K_{\nu+1}(x).
\end{align}
The third contraction of \eqref{eq: partial derivative vertex}, which corresponds to $\Phi_1=A_3$, vanishes in complete analogy to \eqref{eq: vanishing easy contraction with derivative}.

\paragraph{Contribution of fermions in the loop}

The relevant vertices read
\begin{align}\label{eq:FermEffectiveVertex}
\frac{1}{2}\sum_{i=1}^3\tr(\bar \psi_j [G^i]_{jk}[\tilde \phi_i, \psi_k]) +\frac{1}{2}\sum_{i=4}^{3+n_{\scal,\text{easy}}}\tr(\bar \psi_j [G^i]_{jk}[\tilde \phi_i, \gamma_5\psi_k]) + \frac{1}{2}\tr(\bar \psi_j \gamma^\mu[A_\mu, \psi_j])\eqncom
\end{align}
which contribute for $\Phi_1=\scalq_{i,\complicated}$, $\Phi_1=\scalq_{i,\text{easy}}$ and $\Phi_1=A_\mu$, respectively.
The first term gives
\begin{align}
\frac{1}{2}\tr(\contraction{}{\bar \psi}{_j [G^i]_{jk}[\tilde \phi_i, }{\psi}\bar \psi_j [G^i]_{jk}[\tilde \phi_i, \psi_k])
&\simeq \frac{1}{2}[G^i]_{jk} \left(
\langle [\bar{\psi}_j]_{a,n}[\psi_k]_{n^\prime,a}\rangle \tr (E^{n'}{}_n \scalq_i)
-\langle [\bar{\psi}_j]_{n,a}[\psi_k]_{a,n^\prime}\rangle \tr (E^{n}{}_{n'} \scalq_i) \right)
\nonumber\\
&=  N[G^i]_{jk}  [G^l]_{kj} \frac{[t_l]_{n,n'}}{k} \left(\tr K_F^{m=-\frac{k-1}{2}}-\tr K_F^{m=\frac{k+1}{2}}\right) \tr (E^{n}{}_{n'} \scalq_i)\eqncom
\end{align}
where we used the fermionic propagator \eqref{eq:FermionOffDiagProp} and the trace of $K_F$ is with respect to its spinor indices.
Using the anti-commutator relation \eqref{eq:G-anti-commutator} for the $G^i$ matrices, we then find
\begin{equation}
 \begin{aligned}
\frac{1}{2}\tr(\contraction{}{\bar \psi}{_j [G^i]_{jk}[\tilde \phi_i, }{\psi}\bar \psi_j [G^i]_{jk}[\tilde \phi_i, \psi_k])&\simeq
\frac{N}{2k} \tr(\{G^i,G^l\}) (\tr K_F^{m=-\frac{k-1}{2}}-\tr K_F^{m=\frac{k+1}{2}}) \tr (t_l\scalq_i) \\
&=
\frac{N}{k} n_{\psi} (\tr K_F^{m=-\frac{k-1}{2}}-\tr K_F^{m=\frac{k+1}{2}}) \tr (t_i\scalq_i) \eqndot
\end{aligned}
\end{equation}
The evaluation of the second and third term in \eqref{eq:FermEffectiveVertex} is similar to the discussion above, but with $G^i$ replaced by $G^i$ with easy index $i$ and $\gamma^\mu$, respectively. It then follows directly that this contribution vanishes because of the orthogonality of these matrices, cf.\ appendix \ref{app: 10Dto4D}.

\subsection{Total effective vertex}

All vertices come with an overall factor of $\frac{2}{g^2_\YM}$. Adding all the contributions derived above, we arrive at the following total contribution
\begin{align}\label{eq:EffVertex}
V_{\text{eff}}(y)&=  n_{\text{easy}}\frac{2N}{y_3}K^{m^2=\frac{k^2-1}{4}} \tr(\scalq_it_i) \frac{2}{g^2_\YM} \nonumber\\
&\phaneq + n_{\phi,\complicated}\frac{N}{y_3}\left(\frac{k+1}{k}K^{m^2=\frac{(k-2)^2-1}{4}}+\frac{k-1}{k}K^{m^2=\frac{(k+2)^2-1}{4}}\right)\tr(\scalq_i t_i)\frac{2}{g^2_\YM}\nonumber\\
&\phaneq -3(n_{\phi,\complicated}-1)\frac{N}{y_3}\frac{1}{k}\left(K^{m^2=\frac{(k-2)^2-1}{4}}-K^{m^2=\frac{(k+2)^2-1}{4}}\right)\tr(\scalq_i t_i)\frac{2}{g^2_\YM}\nonumber\\
&\phaneq + n_{A,\complicated}\frac{2N}{k}\left(K^{m^2=\frac{(k-2)^2-1}{4}}-K^{m^2=\frac{(k+2)^2-1}{4}}\right) \tr ( t_i \partial_3\scalq_i )\frac{2}{g^2_\YM}\nonumber\\
&\phaneq - n_{A,\complicated}\frac{N}{k}\partial_3\left(K^{m^2=\frac{(k-2)^2-1}{4}}-K^{m^2=\frac{(k+2)^2-1}{4}}\right) \tr ( t_i\scalq_i )\frac{2}{g^2_\YM}\nonumber\\
&\phaneq +n_{\psi}\frac{ N}{k} \left(\tr K_F^{m=-\frac{k-1}{2}}-\tr K_F^{m=\frac{k+1}{2}}\right) \tr (t_i\scalq_i)\frac{2}{g^2_\YM}\eqncom
\end{align}
where all propagators are taken at $y$ and for conciseness we introduced $n_{\text{easy}} = n_{\phi,\text{easy}}+n_{A,\text{easy}}-n_{c}$. 
In particular, the total contribution from all externally contracted fields except for $\Phi_1=\scal_{i,\complicated}$ vanishes.

When contracting the effective vertex \eqref{eq:EffVertex} with a propagator such as in \eqref{eq:LollipopD}, the derivative term can be partially integrated. When we then substitute the dimensional regularised expressions for the propagator from section \ref{dimreg}, the effective vertex becomes
\begin{align}
V_{\text{eff}}(y)= 
\frac{N \tr (t_i\scalq_i)}{16\pi^2 y_3^3}\Biggl[ &
\frac{k^2 (n_{\text{easy}}+n_{\phi,\complicated} -2n_\psi) +n_{\text{easy}} -11n_{\phi,\complicated} -2n_\psi+24n_{A,\complicated} + 12 }{2} \nonumber\\\label{eq:EffVertex2}
&  \times\left\{\frac{1}{\peps}  - \gamma_{\text{E}} + \log(4\pi)+ 2\log(y_3) - 2\Psi(\tfrac{k+1}{2})\right\}\\
& -\frac{k^2 (n_{\text{easy}}+n_{\phi,\complicated} -2n_\psi) +5n_{\phi,\complicated} -3n_{\text{easy}} +6n_\psi-24n_{A,\complicated}}{2}\Biggr]\eqndot\nonumber
\end{align}
We see that the above vanishes exactly when
\begin{align}
\label{eq: conditions on vanishing Veff}
&n_{A,\complicated} = 1,
&& n_{\phi,\complicated} =3,
&& n_{\text{easy}} = 2 n_\psi -3\eqndot
\end{align}
In four dimensions, we have $n_{\text{easy}} \equiv n_{\phi,\text{easy}}+n_{A,\text{easy}}-n_{c} = 3+3-1=5$ and $n_\psi =4$, which satisfies \eqref{eq: conditions on vanishing Veff} such that the effective vertex vanishes. In dimensional regularisation, however, the number of easy gauge fields is $d=3-2\peps$. In dimensional reduction, the number of easy scalars is also  changed in order to preserve supersymmetry, cf.\ the discussion at the end of section \ref{dimreg},  and the total number of easy fields stays five. In other words, the one-loop correction to the vacuum expectation value of all fields vanishes. For the scalar fields, this happens exactly because of supersymmetry. 
It would be interesting to see whether there is a general argument based on supersymmetry that implies that the quantum corrections to (scalar) vevs vanish also at higher loop orders.

\section{Hadamard and zeta-function regularisation \label{Hadamard}}

In this appendix, we summarise the results for  $K(x,x)$ and  $\tr K_F(x,x)$ obtained in section~\ref{dimreg} in the alternative Hadamard as well as zeta-function regularisation,
which are commonly used in AdS. 

\paragraph{Bosonic fields} 

The expression for the scalar loop $K(x,x)$
in zeta-function renormalisation can be found in~\cite{Caldarelli99}, and it reads
\begin{equation}
  K^m(x,x)
    = \frac{g_\YM^2}{2x_3^2}
    \left(-\frac{\frac{1}{3}+m^2}{16\pi^2}+\frac{m^2}{8\pi^2}
      \left[\digamma\left(\nu+\frac{1}{2}\right)-\log\mu\right]\right)\eqndot
  \label{eq:scalar-loop}
\end{equation}
Here, $\mu$ is the renormalisation (mass) scale, and $\digamma$ is  the digamma function.
In~\cite{Kent:2014nya}, $K(x,x)$ is found using Hadamard renormalisation:
\begin{equation}
  K^m(x,x)
    = \frac{g_\YM^2}{2x_3^2}
    \left(-\frac{\frac{1}{3}+m^2}{16\pi^2}+\frac{m^2}{8\pi^2}
      \left[\digamma\left(\nu+\frac{1}{2}\right)-\log\left( \sqrt{2}M \e^{-\gammaE}\right)\right]\right)\eqncom
\end{equation}
where $M$ is the Hadamard renormalisation scale. We notice, as also pointed out
in~\cite{Kent:2014nya}, that the two expressions agree with the identification
\begin{equation}
\mu= \sqrt{2}M \e^{-\gammaE}\eqndot \label{renscale}
\end{equation}

\paragraph{Fermionic fields}
The trace of the fermion loop  in the Hadamard renormalisation scheme can be extracted 
from~\cite{Ambrus:2015mfa}:%
\footnote{There is a misprint in \cite{Ambrus:2015mfa} in the overall sign in the equivalent of  \eqref{eq:fermion-loop}. We thank the authors for communications on this point.
}
\begin{equation}
\begin{aligned}
 \tr K_F^m(x,x)
    &= \frac{g_\YM^2}{2 x_3^3}
    \biggl(\frac{1}{4\pi^2} \Bigl[m^3+m^2+\frac{m}{6}-1\Bigr]
    +\frac{m(m^2-1)}{2\pi^2}
      \left[\digamma\left(m\right)-\log\bigl( \sqrt{2}M \e^{-\gammaE}\bigr)\right]\biggr)\eqndot
      \end{aligned}
  \label{eq:fermion-loop}
\end{equation}
In~\cite{Ambrus:2015mfa}, it is likewise stated (for the stress-energy tensor) that the Hadamard renormalisation for
fermions agrees with the zeta-function one via the identification~(\ref{renscale}). 
However, note that the fermion loop is also calculated using Schwinger-de Witt renormalisation
in \cite{Ambrus:2015mfa}, and this result does \emph{not} match with the Hadamard expression.
Zeta-function renormalisation
for fermions was first carried out in~\cite{Camporesi:1992wn}.
The same remark as made under the discussion of dimensional regularisation 
concerning the chiral rotation of fermions with negative mass applies here.

\paragraph{Implementation} 

For the tadpole diagram, zeta function regularisation gives the same result as dimensional regularisation, presented in (\ref{eq: special tadpole final result}).  However, zeta-function regularisation of the lollipop diagram does not reproduce (\ref{eq: general lollipop diagram final}) but gives a non-vanishing result. 
More precisely, inserting \eqref{eq:scalar-loop} and \eqref{eq:fermion-loop} into the effective vertex \eqref{eq:EffVertex} yields a non-vanishing result, which remains non-vanishing after the contraction with the quantum scalar and the subsequent integration over the vertex position.
The reason for this appears to be that zeta function regularisation 
breaks supersymmetry as observed in other situations \cite{Allen:1983an,Camporesi:1992wn};
recall that supersymmetry in the form of dimensional reduction was crucial for the vanishing of the lollipop diagram in dimensional regularisation.

\bibliographystyle{utphys2}
\bibliography{longpaper}

\end{document}